\documentclass[twocolumn,epjc3]{svjour3}

\usepackage{amsmath,amssymb}

\usepackage{url}
\usepackage{bm}
\usepackage{graphicx}
\usepackage{epsfig}
\usepackage{rotating}
\usepackage{dcolumn}
\usepackage{xcolor}
\usepackage{slashed}
\usepackage{threeparttable}
\usepackage[colorlinks=true
,urlcolor=blue
,anchorcolor=blue
,citecolor=blue
,filecolor=blue
,linkcolor=blue
,menucolor=blue
,pagecolor=blue
,linktocpage=true
,pdfproducer=medialab
]{hyperref}

\RequirePackage{mathptmx}
\RequirePackage{latexsym}
\RequirePackage[numbers,sort&compress]{natbib}

\newcommand{\lsim}{\raisebox{-0.13cm}{~\shortstack{$<$ \\[-0.07cm] $\sim$}}~} 
 
\newcommand{\beq}{\begin{eqnarray}} 
\newcommand{\eeq}{\end{eqnarray}} 
\journalname{Eur. Phys. J. C}

\begin{document}

\title{
\boldmath 
Gluon fusion into Higgs pairs at NLO QCD and the top mass scheme
}


\author{
  J.\,Baglio\thanksref{e1, addr1}
  \and
  F.\,Campanario\thanksref{e2, addr2, addr3}
  \and
  S.\,Glaus\thanksref{e3, addr3, addr4, addr5}
  \and
  M.\,M\"uhlleitner\thanksref{e4, addr3}
  \and
  M.\,Spira\thanksref{e5, addr4}
  \and
\parbox[t]{1.5cm}{
  J.\,Streicher\thanksref{e6, addr1}}
}
\thankstext{e1}{email: \href{mailto:julien.baglio@uni-tuebingen.de}{julien.baglio@uni-tuebingen.de}}
\thankstext{e2}{email: \href{mailto:francisco.campanario@ific.uv.es}{francisco.campanario@ific.uv.es}}
\thankstext{e3}{email: \href{mailto:seraina.glaus@kit.edu}{seraina.glaus@kit.edu}}
\thankstext{e4}{email: \href{mailto:milada.muehlleitner@kit.edu}{milada.muehlleitner@kit.edu}}
\thankstext{e5}{email: \href{mailto:michael.spira@psi.ch}{michael.spira@psi.ch}}
\thankstext{e6}{email: \href{mailto:juraj.streicher@uni-tuebingen.de}{juraj.streicher@uni-tuebingen.de}}


\institute{Institute for Theoretical Physics, Eberhard Karls
  Universit\"{a}t T\"{u}bingen, Auf der Morgenstelle 14, D-72076
  T\"{u}bingen, Germany\label{addr1}
  \and
  Theory Division, IFIC, University of Valencia-CSIC, E-46980 Paterna,
  Valencia, Spain\label{addr2}
  \and
  Institute for Theoretical Physics, Karlsruhe Institute of Technology,
  D-76128 Karlsruhe, Germany\label{addr3}
  \and
  Theory Group LTP, Paul Scherrer Institut, CH-5232 Villigen PSI,
  Switzerland\label{addr4}
  \and
  Institut f\"ur Theoretische Physik, Z\"urich University, CH-8057
  Z\"urich, Switzerland\label{addr5}}

\date{\today}

\maketitle

\begin{abstract} 
We present the calculation of the full next-to-\linebreak leading
order (NLO) QCD corrections to Higgs boson pair production via gluon
fusion at the LHC, including the exact top-mass dependence in the
two-loop virtual and one-loop real corrections. This is the first
independent cross-check of the NLO QCD corrections presented in the
literature before. Our calculation relies on numerical integrations of
Feynman integrals, stabilised with integration-by-parts and a
Richardson extrapolation to the narrow width approximation. We present
results for the total cross section as well as for the invariant
Higgs-pair-mass distribution at the LHC, including for the first time
a study of the uncertainty due to the scheme and scale choice for the
top mass in the loops.
\end{abstract}

\section{Introduction}

Since the discovery of a Higgs boson at the LHC
\cite{Aad:2012tfa,Chatrchyan:2012xdj}, an enormous amount of data
has been collected to analyse the properties of this newly discovered
particle. Up to now the Standard Model (SM) Higgs boson
hypothesis~\cite{Higgs:1964ia,Englert:1964et,Higgs:1964pj,Guralnik:1964eu}
is the most favoured one, even if there is still some room for new physics
effects in the Higgs-coupling measurements. The experimental
determination of the Higgs boson self-coup\-lings, one of the most
important measurements in the Higgs sector and a major goal of the
high-luminosity upgrade of the LHC and of future high-energy
colliders, is still yet to be performed and would give access to the
scalar potential itself which is at the origin of the electroweak
symmetry breaking mechanism. In order to obtain the Higgs self-couplings
and in particular the triple Higgs coupling $\lambda_{H^3}$, the pair
production of Higgs bosons needs to be
considered~{\cite{Dawson:1998py,Djouadi:1999rca}}. For the energies
reachable at the LHC the measurement of the quartic Higgs
self-coupling will be hopeless, since the triple Higgs-production
cross section is too small
\cite{Plehn:2005nk,Binoth:2006ym,Fuks:2015hna}.

The dominant production channel for Higgs pair production at hadron
colliders is gluon fusion. It is mediated by heavy quark-loops at
leading order (LO), in two different topologies: triangle diagrams
containing the triple Higgs coupling, and box diagrams as an irreducible
``background'' for the extraction of
$\lambda_{H^3}$~\cite{Eboli:1987dy,Glover:1987nx,Dicus:1987ic,Plehn:1996wb}.
The NLO QCD corrections have been calculated in
the heavy top-quark limit \linebreak (HTL) some time ago~\cite{Dawson:1998py}, a
framework in which the leading term of a heavy top mass expansion is
obtained so that the NLO calculation reduces to one-loop corrections to
effective $HHg$ and $HHgg$ couplings. The $K$--factor is found to be as
sizeable as the corresponding $K$--factor for single Higgs production,
$K\lsim 2$. The next-to-next-to-leading order (NNLO) QCD corrections in
the same HTL approximation have been computed in
Refs.~\cite{deFlorian:2013uza,deFlorian:2013jea,Grigo:2014jma} for the
total cross section and found to be of the order of $\simeq +20\%$, and
in Ref.~\cite{deFlorian:2016uhr} for the differential distributions.
Soft gluon resummation at next-to-next-to-leading logarithmic (NNLL)
accuracy has been performed \cite{Shao:2013bz,deFlorian:2015moa}, and
the 3-loop matching coefficient has been derived in
Refs.~\cite{Grigo:2014jma,Spira:2016zna}.

There has been an enormous effort in the past few years to reach the
full NLO QCD accuracy in Higgs pair production via gluon fusion
including finite top mass $m_t$ effects. After the calculation of
$m_t$-effects in the real
radiation~\cite{Frederix:2014hta,Maltoni:2014eza} leading to a $-10\%$
reduction of the cross section, a $1/m_t$ expansion up to the order
$\mathcal{O}(1/m_t^{12})$ lead to an estimate of the mass effects at the
order of $\pm 10\%$ for the total cross section at NLO
QCD~\cite{Grigo:2013rya,Grigo:2015dia}. The first calculation of the
full NLO QCD corrections has been performed
in~\cite{Borowka:2016ehy,Borowka:2016ypz} applying numerical methods
based on sector decomposition and contour deformation to master
integrals, and has shown that the $m_t$-effects in the virtual
corrections are of the order of $\sim -5\%$ for the total cross section,
but can amount to $\sim -25\%$ in the tail of the invariant
Higgs-pair-mass ($m_{HH}$) distribution. This fixed-order calculation
has been matched to parton shower programs later~\cite{Heinrich:2017kxx,
Jones:2017giv} and combined with the NNLO QCD corrections in the HTL in
Ref.~\cite{Grazzini:2018bsd}. Up to now no other independent calculation
of the full NLO QCD corrections has been available, but several
approximations have been able to partially reproduce the mass effects
in the virtual
corrections~\cite{Grober:2017uho,Bonciani:2018omm}. Analytic results
in the high-energy limit are also available in
Ref.~\cite{Davies:2018qvx}.

This letter presents the first independent calculation of the
top-mass effects at NLO QCD for Higgs boson pair production via
gluon fusion at the LHC since the original work of
Refs.~\cite{Borowka:2016ehy,Borowka:2016ypz}. Our method is based on
the direct numerical integration of the Feynman integrals of the full
diagrams, using integration-by-parts to improve the stability beyond
the virtual thresholds, defined by intermediate gluon pairs and
top-quark pairs, and a Richardson extrapolation \cite{richardson} to
obtain the final numbers in the narrow-width approximation of the
virtual top quarks. We will present our results for the total cross
section and for the invariant Higgs-pair-mass distribution at a
center-of-mass (c.m.) energy of 14 TeV. We will perform for the first
time an analysis of the uncertainties due to the scheme and scale
chosen for the top mass.

The paper is organised as follows. In Section~\ref{sec:calculation} we
will describe the technical details of our calculation. In
Section~\ref{sec:numresults} we will present our numerical results. We
present the renormalisation and factorisation scale uncertainties in
Section~\ref{sec:scale} and derive the uncertainty due to the top mass in
Section~\ref{sec:topquark}. In Section~\ref{sec:conclusion} we will
close with our conclusions.

\section{Calculation}
\label{sec:calculation}

\subsection{\it Partonic leading order cross section}

At LO, the gluon fusion process is mediated by heavy quark loops. There
are triangle diagrams involving the triple Higgs coupling and box
diagrams with two Yukawa couplings. As the Yukawa coupling is
proportional to the mass of the quarks in the loop, we only consider
the top-quark contribution. Following the conventions of
Ref.~\cite{Dawson:1998py}, the cross section can be decomposed into
form factors after applying two tensor projectors on the matrix
elements, leading to the following LO expression for the partonic
cross section $\hat{\sigma}(gg\to HH)$,
\begin{align}
  \hat{\sigma}_{\rm LO} =
  \frac{G_F^2 \alpha_s^2(\mu_R^2)}{256\left(2\pi\right)^3}
     \int_{\hat{t}_-}^{\hat{t}^+} d\hat{t}\,
  \Big[
  \left|C_\triangle^{} F_\triangle^{} + F_\square^{} \right|^2 +
    \left| G_\square^{} \right|^2\Big],
  \label{eq:xsLO}
\end{align}
where $G_F=1.1663787\cdot 10^{-5}\,\mathrm{GeV^{-2}}$ is the Fermi
constant, $\alpha_s(\mu_R^2)$ is the strong coupling constant
evaluated at the renormalisation scale $\mu_R$, and the Mandelstam
variables $\hat{s}$ and $\hat{t}$ are given by
\begin{align}
  \hat{s}
  & = Q^2 = m_{HH}^2,\nonumber\\
  \hat{t}
  & = -\frac12 \left[ Q^2 - 2 m_H^2 - Q^2\sqrt{1-\frac{4
    m_H^2}{Q^2}}\, \cos\theta\right],
    \label{eq:mandelstam}
\end{align}
with the scattering angle $\theta$ in the partonic c.m. system and
where $m_H$ is the Higgs boson mass. The integrations limits are given
by
\begin{align}
  \hat{t}_\pm = -\frac12 \left[ Q^2 - 2 m_H^2 \mp Q^2\sqrt{1-\frac{4
  m_H^2}{Q^2}}\, \right].
  \label{eq:mandelstam2}
\end{align}
The coefficient $C_\triangle^{}= \lambda_{H^3}v/(Q^2-m_H^2) = 3
m_H^2/(Q^2-m_H^2)$ contains $\lambda_{H^3}$ with $v\simeq 246$~GeV
being the vacuum expectation value of the Higgs field, and the form
factors reduce to $F_\triangle^{} = - F_\square^{} = 2/3$ and
$G_\square^{} = 0$ in the HTL approximation. The full $m_t$-dependence
can be found in Refs.~\cite{Glover:1987nx,Plehn:1996wb}.

\subsection{\it Hadronic cross section}

The NLO QCD corrections include the two-loop virtual corrections to the
triangle and box diagrams, the virtual one-particle-reducible
double-triangle diagrams, and the one-\linebreak loop $2\to 3$ real corrections.
All these contributions are convolved with the parton distributions
functions (PDFs) $f_{i}$ evaluated at the factorisation scale $\mu_F$,
that are included in the parton luminosities $d\mathcal{L}^{ij}/d\tau$,
\begin{align}
  \frac{d\mathcal{L}^{i j}}{d\tau} = \int_\tau^1 \frac{dx}{x}
  f_i\left(x,\mu_F\right) f_j\left(\frac{\tau}{x},\mu_F\right),
  \label{eq:pdflumi}
\end{align}
with $\tau=Q^2/s$, $s$ being the hadronic c.m. energy. The hadronic
cross section can be cast into the form
\begin{align}
  \sigma_{\rm NLO} = \sigma_{\rm LO} + \Delta
  \sigma_{\rm   virt} + \Delta \sigma_{gg} + \Delta \sigma_{qg} + \Delta
  \sigma_{q\bar{q}},
  \label{eq:hadronicxs}
\end{align}
with~\cite{Dawson:1998py}
\begin{align}
  \sigma_{\rm LO}
  & = \int_{\tau_0}^1 d\tau \frac{d\mathcal{L}^{gg}}{d\tau}
    \hat{\sigma}_{\rm LO}\left(Q^2 = \tau s\right),\nonumber\\
  \Delta\sigma_{\rm virt}
  & = \frac{\alpha_s\left(\mu_R^2\right)}{\pi}\int_{\tau_0}^1 d\tau
    \frac{d\mathcal{L}^{gg}}{d\tau} \hat{\sigma}_{\rm LO}\left(Q^2 =
    \tau s\right)\, C_{\rm virt}\left(Q^2\right),\nonumber\\
  \Delta\sigma_{i j}
  & = \frac{\alpha_s\left(\mu_R^2\right)}{\pi}\int_{\tau_0}^1 d\tau
    \frac{d\mathcal{L}^{i j}}{d\tau} \int_{\frac{\tau_0}{\tau}}^1
    \frac{dz}{z}\, \hat{\sigma}_{\rm LO}\left(Q^2 = z\tau s\right)
    C_{i j}(z),
    \label{eq:nlocorrections}
\end{align}
for $ij = gg$, $\displaystyle\sum_{q,\bar{q}} qg$, and
$\displaystyle\sum_q q\bar{q}$, $z=Q^2/\tau s$, and $\tau_0 = 4
m_H^2/s$. We include five external massless quark flavours. The coefficients $C_{virt}$ of the virtual and $C_{ij}$ of the
real corrections in the HTL have been obtained in
Ref.~\cite{Dawson:1998py} and are given by

\begin{align}
 C_{virt} & = \frac{11}{2} + \pi^2 + C^\infty_{\triangle\triangle} +
\frac{33-2N_F}{6} \log\frac{\mu_R^2}{Q^2}, \nonumber \\
 C_{\triangle\triangle} & = 
\Re e~\frac{\int_{\hat t_-}^{\hat t_+} d\hat t \left\{ c_1 \left[
(C_\triangle F_\triangle + F_\Box) + \frac{p_T^2}{\hat t}
G_\Box \right] + (\hat t \leftrightarrow \hat u) \right\}}
{\int_{\hat t_-}^{\hat t_+} d\hat t \left\{ |C_\triangle F_\triangle +
F_\Box |^2 + |G_\Box|^2 \right\}}, \nonumber \\
C^\infty_{\triangle\triangle} & = \left. C_{\triangle\triangle}
\right|_{c_1 = 2/9}, \nonumber \\
C_{gg} & = -z P_{gg}(z) \log\frac{\mu_F^2}{\tau s} - \frac{11}{2}
(1-z)^3 \nonumber \\
& \qquad + 6[1+z^4+(1-z)^4] \left(\frac{\log(1-z)}{1-z}\right)_+, \nonumber \\
C_{gq} & = -\frac{z}{2} P_{gq}(z) \log\frac{\mu_F^2}{\tau s (1-z)^2} +
\frac{2}{3} z^2 - (1-z)^2, \nonumber \\
C_{q\bar q} & = \frac{32}{27} (1-z)^3,
    \label{eq:coeffvirt}
\end{align}
where $C^\infty_{\triangle\triangle}$ denotes the contribution of the
one-particle reducible diagrams (see Fig.~\ref{fg:virtdia}b) in the HTL
with the transverse momentum $p_T^2 = (\hat t\hat u - m_H^4)/Q^2$
involving $\hat u = 2m_H^2 - Q^2 - \hat t$. $P_{gg}(z)$ and $P_{gq}(z)$
are the corresponding Altarelli-Parisi splitting functions
\cite{Altarelli:1977zs}, given by
\begin{align}
P_{gg}(z) &= 6\left\{ \left(\frac{1}{1-z}\right)_+
+\frac{1}{z}-2+z(1-z) \right\}
\nonumber \\
& \qquad\qquad\qquad\qquad\qquad\qquad + \frac{33-2N_F}{6}\delta(1-z), 
\nonumber \\
P_{gq}(z) &= \frac{4}{3} \frac{1+(1-z)^2}{z},
\end{align}
with $N_F=5$ in our calculation.  For the LO cross section
$\hat{\sigma}_{\rm LO}(Q^2)$ the full quark-mass dependence is taken
into account at the integrand-level.  These expressions can easily be
converted into the differential cross section with respect to the
invariant Higgs-pair mass,
\begin{align}
  \frac{d\sigma_{\rm LO}}{dQ^2}
  & = \left. \frac{d\mathcal{L}^{gg}}{d\tau}
    \frac{\hat{\sigma}_{\rm LO}\left(Q^2\right)}{s} \right|_{\tau =
\frac{Q^2}{s}},\nonumber\\
  \frac{d\Delta\sigma_{\rm virt}}{dQ^2}
  & = \left. \frac{\alpha_s\left(\mu_R^2\right)}{\pi}
    \frac{d\mathcal{L}^{gg}}{d\tau} \frac{\hat{\sigma}_{\rm LO}\left(Q^2
    \right)}{s}\, C_{\rm virt}\left(Q^2\right) \right|_{\tau = \frac{Q^2}{s}},
    \nonumber\\
  \frac{d\Delta\sigma_{i j}}{dQ^2}
  & = \left. \frac{\alpha_s\left(\mu_R^2\right)}{\pi}\int_{\frac{Q^2}{s}}^1
    \frac{dz}{z^2} \frac{d\mathcal{L}^{i j}}{d\tau}
    \, \frac{\hat{\sigma}_{\rm LO}\left(Q^2\right)}{s}
    C_{i j}(z) \right|_{\tau = \frac{Q^2}{zs}}.
    \label{eq:nlodiff}
\end{align}

\subsection{\it Virtual corrections}

\begin{figure*}[hbt]

 \centering
  \includegraphics[scale=1.00]{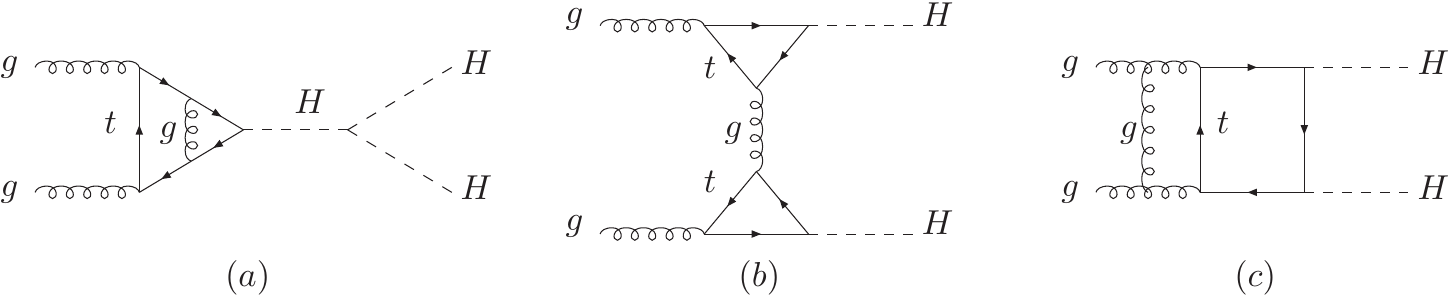}

  \caption[]{\it Typical two-loop diagrams
contributing to Higgs-boson pair production via gluon fusion: (a)
triangle diagram, (b) one-particle reducible diagram, (c) box diagram.}
\label{fg:virtdia} \end{figure*}
The two-loop virtual corrections split into three different types of
contributions: triangle diagrams with a virtual Higgs propagator in the
s-channel, box diagrams, and one-particle reducible diagrams. Typical
examples of these three classes are shown in Fig.~\ref{fg:virtdia}.
According to Eq.~(\ref{eq:nlodiff}) the two-loop triangle diagrams
coincide with the production of a single Higgs boson with virtuality
$Q^2$ so that they can be taken from this NLO calculation
\cite{Graudenz:1992pv, Spira:1995rr, Harlander:2005rq,
Anastasiou:2009kn, Aglietti:2006tp}.  On the other hand the one-particle
reducible diagrams involving two triangle loops of the Higgs coupling to
one on-shell and one off-shell gluon can be constructed from the LO
(one-loop) calculation of the Higgs decay $H\to Z\gamma$
\cite{Cahn:1978nz,Bergstrom:1985hp} where the couplings and colour
factors have to be adjusted appropriately and the $Z$ mass has to be
replaced by the virtuality of the gluon in the t/u-channel. This results
in the one-particle reducible contribution to the virtual corrections
\begin{align}
 c_1 & = 2 \Big[ I_1(\tau,\lambda_{\hat t})
                -I_2(\tau,\lambda_{\hat t}) \Big]^2, \nonumber \\
I_1(\tau,\lambda) & = \frac{\tau\lambda}{2(\tau-\lambda)} +
\frac{\tau^2\lambda^2}{2(\tau-\lambda)^2} \left[ f(\tau) - f(\lambda)
\right] \nonumber \\
& \qquad\qquad + \frac{\tau^2\lambda}{(\tau-\lambda)^2} \left[ g(\tau) -
g(\lambda) \right], \nonumber \\
I_2(\tau,\lambda) & = - \frac{\tau\lambda}{2(\tau-\lambda)}\left[
f(\tau)
- f(\lambda) \right],
\end{align}
with $\tau = 4m_t^2/m_H^2$, $\lambda_{\hat t} = 4 m_t^2/\hat t$ and the
functions
\begin{align}
f(\tau) & = \left\{ \begin{array}{ll}
\displaystyle \arcsin^2 \frac{1}{\sqrt{\tau}} & \tau \ge 1 \\
\displaystyle - \frac{1}{4} \left[ \log \frac{1+\sqrt{1-\tau}}
{1-\sqrt{1-\tau}} - i\pi \right]^2 & \tau < 1
\end{array} \right. \, , \nonumber \\
g(\tau) & = \left\{ \begin{array}{ll}
\displaystyle \sqrt{\tau-1} \arcsin \frac{1}{\sqrt{\tau}} & \tau \ge 1
\\
\displaystyle \frac{\sqrt{1-\tau}}{2} \left[ \log \frac{1+\sqrt{1-\tau}}
{1-\sqrt{1-\tau}} - i\pi \right] & \tau < 1
\end{array} \right. \, .
\end{align}
This expression has to be inserted in the $C_{\triangle\triangle}$
coefficient of Eq.~(\ref{eq:coeffvirt}) and agrees with the previous
analytical result of Ref.~\cite{Degrassi:2016vss}.

The cumbersome part of this calculation is the computation of the
two-loop box diagrams. We have performed a Feynman parametrisation of
the full virtual diagrams individually without any tensor reduction to
master integrals. The ultraviolet singularities are extracted by
appropriate end-point subtractions. Special care, however, is required
for the infrared divergent diagrams of the type of e.g.
Fig.~\ref{fg:virtdia}c that correspond to gluon-rescattering involving a
threshold starting at $Q^2=0$, i.e.~in the whole kinematical range of
the process. The Feynman parametrisation can be adjusted such that the
(singular) denominator is expressed as a second-order polynomial in
terms of one of the Feynman parameters (here $x_6$, while $x_1,\ldots,x_5$
correspond to the other parameters),
\begin{align}
 I & = \int d^5 \vec x dx_6 \frac{x_6^{1+\epsilon} H(\vec x,x_6)}{N(\vec
x,x_6)^{3+2\epsilon}} \qquad [\vec x = (x_1,\ldots,x_5)], \nonumber \\
N(\vec x,x_6) & = a x_6^2 + b x_6 + c,
  \label{eq:irint}
\end{align}
where $H(\vec x,x_6)$ denotes the numerator related to the full spinorial
structure of the diagram and
\begin{align}
a,c = {\cal O} \left(\frac{1}{m_t^2}\right), \qquad b = 1 + {\cal O}
\left( \frac{1}{m_t^2} \right) \, ,
\end{align}
where the coefficient $c$ can be expressed as a pure product of a
kinematical ratio of ${\cal O} (1/m_t^2)$ and other Feynman parameters.
Based on the fact that the relative infrared singularities are universal
we constructed a subtraction term by keeping only $b$ and $c$ in the
denominator $N_0(\vec x,x_6) =  b x_6 + c$ thus rewriting the integral
of Eq.~(\ref{eq:irint}) as
\begin{align}
 I & = I_1 + I_2, \nonumber \\
 I_1 & = \int d^5 \vec x dx_6 \left\{
\frac{x_6^{1+\epsilon} H(\vec x,x_6)}{N(\vec x,x_6)^{3+2\epsilon}}
- \frac{x_6^{1+\epsilon} H(\vec x,0)}{N_0(\vec x,x_6)^{3+2\epsilon}} \right\},
\nonumber \\
 I_2 & = \int d^5 \vec x dx_6
\frac{x_6^{1+\epsilon} H(\vec x,0)}{N_0(\vec x,x_6)^{3+2\epsilon}}\, .
\end{align}
The integral $I_1$ can now be expanded in $\epsilon$ leading to
numerically finite integrals for each expansion coefficient, while the
integral $I_2$ can be integrated over $x_6$ yielding hypergeometric
functions. The transformation properties of the latter (related to an
inversion of the kinematical argument) allow for a clean separation of
the infrared singularities. The remaining singularities in the
coefficient $c$ can be treated by end-point subtractions. The
cumbersome treatment of these diagrams can be related to the
appearance of two different scales, $m_t$ and $Q$, that control the
high-scale and low-scale parts of the whole calculation in the sense
of an effective theory, the HTL, where the high scale is given by the
top mass and the low scale by $Q$. This method of constructing the
infrared subtraction term has been developed within the old NLO
calculation of single Higgs production \cite{Graudenz:1992pv,
  Spira:1995rr} for internal numerical checks. Since the integration
over $\hat t$ is not finite for the individual two-loop diagrams we
have introduced a cut at the integration bounds as well as a
suitable substitution to stabilise the integration close to the
bounds. By varying the cut around its central value we have checked
that the total sum of all two-loop box diagrams is finite and
independent of the particular value of this cut.

For the analytical continuation of our virtual amplitudes above the
thresholds we introduced a small imaginary part of the virtual top mass,
$m_t^2 \to m_t^2 (1-i\bar\epsilon)$ in our numerical integration.
However, to stabilise the numerical integration above the thresholds we
had to perform integration by parts in one of the involved Feynman
parameters in order to reduce the power of the corresponding
denominator for each diagram individually. In this way we achieved a
reliable accuracy of our numerical integrations for $\bar\epsilon$ values
down to about 0.1. In order to obtain the result in the narrow-width
approximation ($\bar\epsilon \to 0$) we performed a Richardson
extrapolation \cite{richardson} applied to the results for different
values of $\bar\epsilon$\footnote{It should be noted that a Richardson
extrapolation of the integrand {\it before} integration provides an
alternative method to stabilise the numerical integration.}.
In particular, we use $\bar\epsilon$ in the range given by
$\bar\epsilon_n= 0.05\times 2^n$, with $n=0,\ldots,9$. In the dominant
region, we use the set $n=1,\dots,9$, with the exception of the bins
in the range $Q\in [300-475]$ GeV where the complete set of values is
used. Starting at $Q=700$ GeV, we restrict ourselves to values in the
range $n=1,\ldots,5$. This allows us to obtain a series of
extrapolated results up to the ninth order in the dominant region and
up to the fifth order in the tails. We define a theoretical
error estimate due to the Richardson extrapolation as the difference
of the fifth and the fourth order extrapolated results. Moreover, this
error is multiplied by a factor of two close to the top threshold, in
order to be conservative. The obtained estimated Richardson
extrapolated error falls below the percent level of accuracy and is
added in quadrature to the statistical integrated error.

The top mass has been renormalised in the on-shell \linebreak scheme and the strong
coupling constant in the $\overline{\rm MS}$ scheme with 5 active flavours,
i.e.~with a decoupled top quark. We have achieved a finite result for the
virtual corrections by subtracting the virtual correction in the HTL so
that our numerical integration yields the NLO mass effects only. The
virtual corrections in the HTL have then been added back by the results
of {\tt HPAIR}\footnote{The program can be downloaded at
\href{http://tiger.web.psi.ch/hpair/}{http://tiger.web.psi.ch/hpair/}.}.
The calculation of each two-loop box diagram has been performed
independently at least twice with different Feynman parametrisations.

\subsection{\it Real corrections}

We are left with the calculation of the coefficients $C_{ij}$, or more
specifically the calculation of the finite $m_t$-effects in the real
corrections,  $\Delta \sigma_{ij}^{\rm mass} = \Delta \sigma_{ij} -
\Delta \sigma_{ij}^{\rm HTL}$, where $\Delta \sigma_{ij}^{\rm HTL}$
involves the full mass-dependent LO contribution as exemplified in
Eqs.~(\ref{eq:nlocorrections}, \ref{eq:nlodiff}). The HTL expressions
for the coefficients $C_{ij}$ are given in
Eq.~(\ref{eq:coeffvirt})~\cite{Dawson:1998py} and can be calculated using
the program {\tt HPAIR}.

To obtain the mass effects, we use the fact that the infrared
divergences in the real corrections are universal and are the same in
the full calculation and in the HTL approximation. At any given
phase-space point we can subtract the HTL result from the full
calculation, obtaining an infrared-finite result which is exactly the
remainder due to the mass effects in the full real corrections,
\begin{align}
 d\Delta\hat{\sigma}_{ij}^{\rm mass} = d\Delta\hat{\sigma}_{ij} -
  d\hat{\sigma}_{\rm LO}(\tilde p_i) \frac{d\Delta\hat{\sigma}_{ij}^{\rm
  HTL}(p_i)}{d\hat{\sigma}_{\rm LO}^{\rm HTL}(\tilde p_i)}.
\end{align}
In order to calculate the LO matrix elements we need to map the full
$2\to 3$ phase-space onto an on-shell LO $2\to2$ sub-space, leading to a
transformation of the 4-momenta $p_i \to \tilde p_i$. For this
purpose, we use the mapping of Ref.~\cite{Catani:1996vz} for the case
of initial-state emitter and initial-state spectator to build our local
subtraction term. The HTL matrix elements are calculated analytically.
In order to achieve a good numerical stability we have implemented a
technical collinear cut in the scattering angle integration of the extra
parton in the final state, $|\cos\theta| < 1 - \delta$, with $\delta =
10^{-4}$. It has been checked that this value does not affect the
physical results by varying $\delta$ around our nominal choice.

The full one-loop matrix elements contain triangle, box, and pentagon
diagrams. They are generated using {\tt Feyn-\linebreak Arts}
\cite{Hahn:2000kx} and {\tt FormCalc}~\cite{Hahn:1998yk}, while the
scalar one-loop integrals are numerically calculated using the library
{\tt COLLIER 1.2}~\cite{Denner:2016kdg} interfaced to the output of {\tt
FormCalc} with an in-house library. We have performed two independent
calculations with different parametrisations of the $2\to3$ phase-space
and different versions of {\tt FeynArts} and {\tt FormCalc}. We have
also cross-checked the mass effects of the real corrections against
Refs.~\cite{Frederix:2014hta,Maltoni:2014eza,Borowka:2016ehy,Borowka:2016ypz}
and we have obtained mutual agreement.

\section{Numerical results}
\label{sec:numresults}

\begin{figure*}[t!]
  \centering
  \includegraphics[scale=0.63]{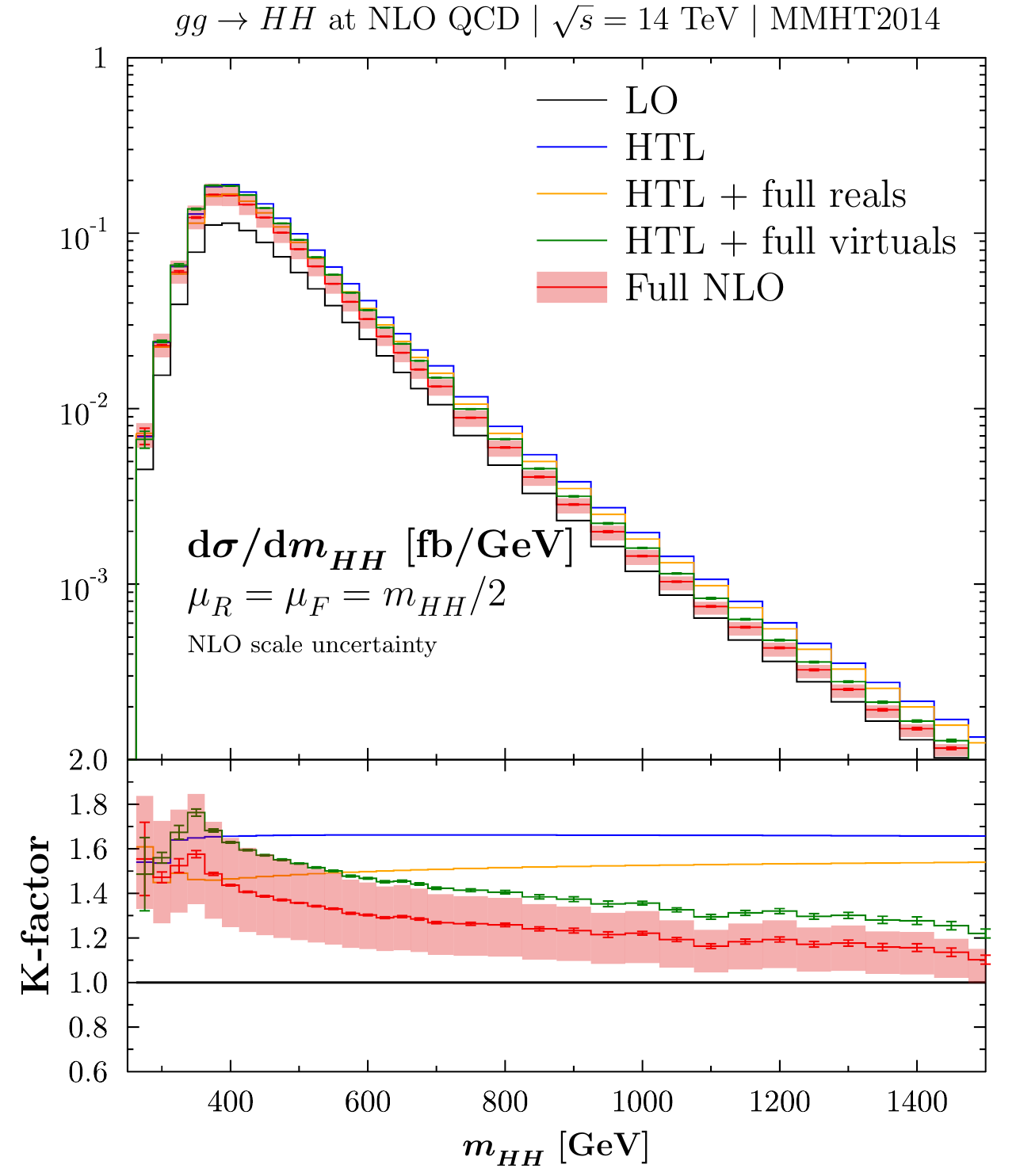}
  \hspace{3mm}
  \includegraphics[scale=0.63]{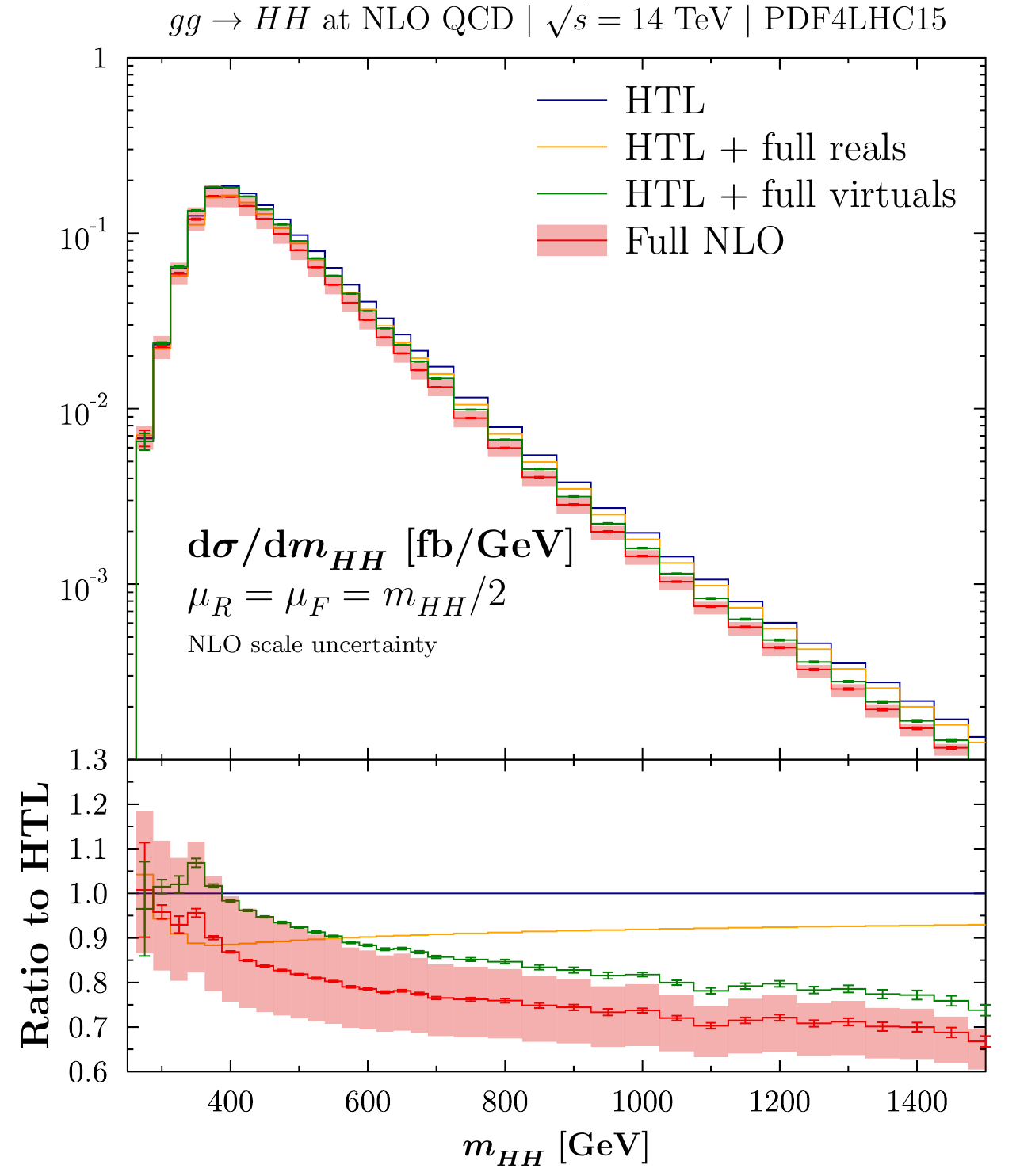}
  \caption[]{Invariant Higgs-pair-mass distributions for Higgs boson
    pair production via gluon fusion at the 14 TeV LHC as a function
    of $Q=m_{HH}$. LO results (in black), HTL results (in blue), HTL
    results including the full real corrections (in yellow), HTL
    results including the full virtual corrections (in green,
    including the numerical error), and the full NLO QCD results
    (in red, including the numerical error). Left: Results with the
    {\tt MMHT2014} PDF set, the insert below displays the $K$-factors
    for the different results. Right: Results with the {\tt
      PDF4LHC15} PDF set, the insert below displays the ratio to the
    NLO HTL result for the different calculations. The red band
    indicates the renormalisation and factorisation scale
    uncertainties for results including the full NLO QCD corrections.}
  \label{fig:distrib}
\end{figure*}

We present our numerical results at the LHC for a c.m.~energy of
$\sqrt{s}=14$~TeV. We use $m_H = 125$~GeV and $m_t=172.5$~GeV. We have
performed the calculation using two NLO PDF sets, {\tt
MMHT2014}~\cite{Harland-Lang:2014zoa} and {\tt
PDF4LHC15}~\cite{Butterworth:2015oua} as implemented in the {\tt
LHAPDF-6} library~\cite{Buckley:2014ana}. Our central scale choice is
$\mu_R=\mu_F=\mu_0 = m_{HH}/2$, and $\alpha_s(M_Z^2)$ is set according
to the PDF set chosen, with a NLO running in the five-flavour scheme.
We recall that $Q$ and $m_{HH}$ refer to the same physical quantity. We
have performed the whole calculation twice independently using also
different parametrisations of the virtual Feynman integrals and of the
real phase-space. Both independent calculations agree with each other
within the numerical errors. Note that our calculation has been
performed in the narrow-width approximation for the top
quark. The finite-width effect amounts to $\sim -2\%$ at LO for the
total cross section, corresponding to a maximum deviation of $\sim
-4\%$ at the $t\bar{t}$ threshold in the invariant Higgs-pair-mass
distribution~\cite{Maltoni:2014eza}.

We have calculated a grid of $Q$-values from $Q=250~\mathrm{GeV}$ to
$Q=1500~\mathrm{GeV}$, so that we obtain the invariant Higgs-pair-mass
distribution depicted in Fig.~\ref{fig:distrib}. We compare our full
results, displayed in red, to three different approximations: The HTL
results according to Ref.~\cite{Dawson:1998py} are displayed in blue;
the HTL plus the mass effects in the real corrections only are shown in
yellow; the HTL plus the mass effects in the virtual corrections only
are presented in green. The error bars due to the total
  numerical errors are also given for the green and red
  histograms. They are determined from the individual integration
  errors and the errors due to the Richardson extrapolation -- added in
  quadrature. The numerical errors are negligible for the other
  predictions (HTL and HTL plus the mass effects in the real
  corrections only). The red band indicates the renormalisation and
  factorisation scale uncertainties of our
  prediction for the full NLO QCD predictions, see next section for
  more details.
In the case of the {\tt MMHT2014} PDF set we also compare to
the LO results. We reproduce the results of
Refs.~\cite{Frederix:2014hta,Maltoni:2014eza} for the mass effects in
the real corrections, which are mildly varying from $m_{HH} = 400$~GeV
to $m_{HH} = 1500$~GeV, and are of the order of $-10\%$. The global
$K$-factor displayed on the left-hand-side of Fig.~\ref{fig:distrib} is
decreasing in the whole range and gets close to 1 for large $m_{HH}$
values. The mass effects in the virtual corrections reach $\simeq -20\%$
at large $m_{HH}$ values, in agreement with the findings of
Refs.~\cite{Borowka:2016ehy,Borowka:2016ypz}. The mass effects are
negative in accordance with the expected restoration of partial-wave
unitarity in the high-energy limit.

From the differential distribution we calculate the total cross section
using a numerical integration over $Q$ with the trapezoidal method
supplemented by a Richardson extrapolation \cite{richardson} for $Q>300$
GeV, while for $Q<300$ GeV we have used the extension of Boole's rule to
six nodes \cite{Abramowitz}. We obtain the following results for the NLO
cross sections (the numbers in parenthesis
indicate the numerical errors, i.e. the quadratic sum of the
statistical and the Richardson extrapolation error),
\begin{align}
  \sigma_{gg\to HH}^{\rm PDF4LHC} = 32.78(7)\,\mathrm{fb}, \quad
  \sigma_{gg\to HH}^{\rm MMHT} = 33.33(7)\,\mathrm{fb}.
  \label{eq:finalxs}
\end{align}
These have to be compared to the HTL results,
\begin{align}
  \sigma_{gg\to HH}^{\rm HTL, PDF4LHC} = 38.66\,\mathrm{fb}, \quad
  \sigma_{gg\to HH}^{\rm HTL, MMHT} = 39.34\,\mathrm{fb}.
  \label{eq:finalxshql}
\end{align}
Comparing Eq.~(\ref{eq:finalxs}) and Eq.~(\ref{eq:finalxshql}), we derive
$\simeq -15\%$ top-mass effects at NLO on the total cross
section. We are in mutual agreement with the results of
Refs.~\cite{Borowka:2016ehy,Borowka:2016ypz} within the numerical
uncertainties and taking into account the small difference in the
choice of the top mass value.

\section{Factorisation and renormalisation scale dependence}
\label{sec:scale}

For the estimate of the residual theoretical uncertainties of our NLO
results we derive the uncertainties due to the \linebreak choices of the
factorisation and renormalisation scales by varying both by a factor of
two up and down around the central scale choice $\mu_R=\mu_F=m_{HH}/2$,
but avoid a splitting of both scales by more than a factor of two. This
leads to seven points in the $\mu_R-\mu_F$-plane, since both scale
dependences are monotonic. The scale uncertainty is then determined by
the maximal and minimal cross section values around the central values.
The factorisation scale enters the luminosity factors and the HTL parts
of the coefficients $C_{ij}$ in Eq.~(\ref{eq:nlodiff}) so that the
numerical two-loop results of the NLO mass effects are not affected. The
same is true for the renormalisation scale dependence, since the strong
coupling constant enters only as a universal factor in each perturbative
order and the explicit renormalisation scale appears only in the
HTL-part of the virtual corrections. In total we find the following
scale dependences for the differential cross section for four distinct
values of $Q$,
\begin{align}
\frac{d\sigma(gg\to HH)}{dQ}\Big|_{Q=300~{\rm GeV}} & = 0.0298(7)^{+15.3\%}_{-13.0\%}\,
\mathrm{fb/GeV},\nonumber\\
\frac{d\sigma(gg\to HH)}{dQ}\Big|_{Q=400~{\rm GeV}} & = 0.1609(4)^{+14.4\%}_{-12.8\%}\,
\mathrm{fb/GeV},\nonumber\\
\frac{d\sigma(gg\to HH)}{dQ}\Big|_{Q=600~{\rm GeV}} & = 0.03204(9)^{+10.9\%}_{-11.5\%}\,
\mathrm{fb/GeV},\nonumber\\
\frac{d\sigma(gg\to HH)}{dQ}\Big|_{Q=1200~{\rm GeV}} & = 0.000435(4)^{+7.1\%}_{-10.6\%}\,
\mathrm{fb/GeV},
\end{align}
for PDF4LHC parton densities,
while the total cross section develops the uncertainties
\begin{align}
\sigma(gg\to HH) = 32.78(7)^{+13.5\%}_{-12.5\%}\,\mathrm{fb}.
\end{align}
The last result is in mutual agreement with Refs.~\cite{Borowka:2016ehy,
Borowka:2016ypz} within the numerical uncertainties and taking into
account the small difference in the choice of the top mass value.

\section{Uncertainties due to the top mass scheme}
\label{sec:topquark}

For the uncertainty related to the scheme and scale choice of the top
mass we have calculated the total NLO results for the differential
gluon-fusion cross section for the $\overline{\rm MS}$ top mass at
different scale choices and have compared to our default prediction
using the top quark pole mass both in the loop propagators and in the
Yukawa coupling. We have used an N$^3$LO evolution and
conversion of the pole into the $\overline{\rm MS}$ mass at the input
scale given by the $\overline{\rm MS}$ top mass itself. This leads, for
our choice of $m_t=172.5$ GeV for the top pole mass to an $\overline{\rm
MS}$ mass of $\overline{m}_t(\overline{m}_t) = 163.02$ GeV. The
renormalisation of the top mass has been adjusted accordingly in our
calculation, and we have switched to an $\overline{\rm MS}$ mass both
in the loop propagators and in the Yukawa coupling.
We present the top-quark scheme uncertainties at four selected values
of $Q$ in the invariant Higgs-pair mass differential cross section. We
take the maximum and minimum differential cross sections when the
scale of the $\overline{\rm MS}$ top quark mass is varied in the range
$Q/4$ and $Q$, compared to our default pole mass predictions, and we
obtain the following variations,
\begin{align}
\frac{d\sigma(gg\to HH)}{dQ}\Big|_{Q=300~{\rm GeV}} & = 0.0298(7)^{+6\%}_{-34\%}\,
\mathrm{fb/GeV},\nonumber\\
\frac{d\sigma(gg\to HH)}{dQ}\Big|_{Q=400~{\rm GeV}} & =  0.1609(4)^{+0\%}_{-13\%}\,
\mathrm{fb/GeV},\nonumber\\
\frac{d\sigma(gg\to HH)}{dQ}\Big|_{Q=600~{\rm GeV}} & = 0.03204(9)^{+0\%}_{-30\%}\,
\mathrm{fb/GeV},\nonumber\\
\frac{d\sigma(gg\to HH)}{dQ}\Big|_{Q=1200~{\rm GeV}} & = 0.000435(4)^{+0\%}_{-35\%}\,
\mathrm{fb/GeV},
\end{align}
using PDF4LHC parton densities. The top-quark scheme uncertainty is
significant over the whole range of $m_{HH}$.
Note that a similar result has been observed in single Higgs
production for large Higgs masses which correspond to our triangle
diagram involving the triple Higgs coupling. Furthermore, this scheme
uncertainty is reduced by roughly a factor of two from LO to NLO.
The prediction involving the top pole mass, that we take as our central
prediction, is the maximal prediction for high $m_{HH}$ values. The
uncertainties induced by the top-mass scheme and scale choice on the
total cross section at NLO will be given in a forthcoming
publication~\cite{newpaper}.

\section{Conclusions}
\label{sec:conclusion}

We have presented the calculation of the full NLO QCD corrections to
Higgs-boson pair production via gluon fusion for the top-loop
contributions. This has been performed by numerical integrations of
the involved virtual two-loop corrections to the four-point functions,
while the results of the single-Higgs case have been translated to the
three-point contributions that involve the trilinear Higgs
self-coupling. The one-particle reducible contributions that appear
for the first time at NLO have been inferred from the explicit
analytical one-loop results for $H\to Z\gamma$, where the $Z$-boson
mass plays the role of the virtuality of the gluon in the dressed
$Hgg^\ast$ vertex. In order to isolate the ultraviolet, infrared and
collinear divergences, we have performed appropriate end-point
subtractions at the integrand level and described the explicit
construction of infrared subtraction terms that allow for a clean
separation of the infrared singularities from the regular rest. The
real corrections have been obtained by generating the full matrix
elements with automatic tools. We have constructed the infrared and
collinear subtraction term as the heavy-top limit of the real matrix
elements involving the fully massive LO sub-matrix element. Adding
back the full results in the heavy-top limit completed the full real
corrections. The final results we have obtained agree with previous
calculations for the individual finite parts of the real and virtual
corrections. We find finite NLO mass effects that are up to $-30\%$
for large invariant Higgs-pair masses, while the total NLO top-mass
effects modify the total cross section by about $-15\%$.

We have studied the theoretical uncertainties related to variations of
the renormalisation and factorisation scales and have found agreement
with the previously known results finding uncertainties at the level
of $10-15\%$. A novel outcome of our calculation is the additional
uncertainty induced by the scheme and scale dependence of the top mass
that can be significant, amounting to $+6\%/-34\%$ at $m_{HH}=300$~GeV
and $+0\%/-35\%$ at $m_{HH} = 1200$~GeV. The induced uncertainty on
the total cross section will be given in a forthcoming
publication~\cite{newpaper}.

In the future we plan to extend our calculation to beyond-the-SM
models as e.g.~the 2HDM or MSSM.
\bigskip

\begin{acknowledgements}
We are grateful to S.~Dittmaier for providing us with a copy of his old
{\tt mathematica} programs for the QCD corrections in the HTL related
to Ref.~\cite{Dawson:1998py}.
J.~B. and J.~S. acknowledge the support from the Institutional Strategy
of the University of T\"ubingen (DFG, ZUK 63), the DFG Grant JA 1954/1,
and the Carl Zeiss Foundation. F.~C. acknowledges financial support by
the Generalitat Valenciana, Spanish Government and ERDF funds from the
European Commission (Grants No. RYC-2014-16061, SEJI-2017/2017/019,
FPA2017-84543-P,FPA2017-84445-P, and SEV-2014-0398).
The work of S.~G. is supported by the Swiss National Science
Foundation (SNF). This work has been performed thanks to the support
of the state of Baden-W\"urttemberg through bwHPC and the German
Research Foundation (DFG) through grant no INST 39/963-1 FUGG.
\end{acknowledgements}

{\small
\bibliographystyle{apsrev4-1}
\bibliography{hh_letter}

\begin{thebibliography}{53}%
\makeatletter
\providecommand \@ifxundefined [1]{%
 \@ifx{#1\undefined}
}%
\providecommand \@ifnum [1]{%
 \ifnum #1\expandafter \@firstoftwo
 \else \expandafter \@secondoftwo
 \fi
}%
\providecommand \@ifx [1]{%
 \ifx #1\expandafter \@firstoftwo
 \else \expandafter \@secondoftwo
 \fi
}%
\providecommand \natexlab [1]{#1}%
\providecommand \enquote  [1]{``#1''}%
\providecommand \bibnamefont  [1]{#1}%
\providecommand \bibfnamefont [1]{#1}%
\providecommand \citenamefont [1]{#1}%
\providecommand \href@noop [0]{\@secondoftwo}%
\providecommand \href [0]{\begingroup \@sanitize@url \@href}%
\providecommand \@href[1]{\@@startlink{#1}\@@href}%
\providecommand \@@href[1]{\endgroup#1\@@endlink}%
\providecommand \@sanitize@url [0]{\catcode `\\12\catcode `\$12\catcode
  `\&12\catcode `\#12\catcode `\^12\catcode `\_12\catcode `\%12\relax}%
\providecommand \@@startlink[1]{}%
\providecommand \@@endlink[0]{}%
\providecommand \url  [0]{\begingroup\@sanitize@url \@url }%
\providecommand \@url [1]{\endgroup\@href {#1}{\urlprefix }}%
\providecommand \urlprefix  [0]{URL }%
\providecommand \Eprint [0]{\href }%
\@ifxundefined \urlstyle {%
  \providecommand \doi  [0]{\begingroup \@sanitize@url \@doi}%
  \providecommand \@doi [1]{\endgroup \@@startlink {\doibase
  #1}doi:\discretionary {}{}{}#1\@@endlink }%
}{%
  \providecommand \doi  [0]{doi:\discretionary{}{}{}\begingroup
  \urlstyle{rm}\Url }%
}%
\providecommand \doibase [0]{http://dx.doi.org/}%
\providecommand \Doi [0]{\begingroup \@sanitize@url \@Doi }%
\providecommand \@Doi  [1]{\endgroup\@@startlink{\doibase#1}\@@Doi}%
\providecommand \@@Doi [1]{#1\@@endlink}%
\providecommand \selectlanguage [0]{\@gobble}%
\providecommand \bibinfo  [0]{\@secondoftwo}%
\providecommand \bibfield  [0]{\@secondoftwo}%
\providecommand \translation [1]{[#1]}%
\providecommand \BibitemOpen [0]{}%
\providecommand \bibitemStop [0]{}%
\providecommand \bibitemNoStop [0]{.\EOS\space}%
\providecommand \EOS [0]{\spacefactor3000\relax}%
\providecommand \BibitemShut  [1]{\csname bibitem#1\endcsname}%
\bibitem [{\citenamefont {Aad}\ \emph {et~al.}(2012)\citenamefont {Aad} \emph
  {et~al.}}]{Aad:2012tfa}%
  \BibitemOpen
  \bibfield  {author} {\bibinfo {author} {\bibfnamefont {G.}~\bibnamefont
  {Aad}} \emph {et~al.} (\bibinfo {collaboration} {ATLAS}),\ }\Doi
  {10.1016/j.physletb.2012.08.020} {\bibfield  {journal} {\bibinfo  {journal}
  {Phys. Lett.},\ }\textbf {\bibinfo {volume} {B716}},\ \bibinfo {pages} {1}
  (\bibinfo {year} {2012})},\ \Eprint {http://arxiv.org/abs/1207.7214}
  {arXiv:1207.7214 [hep-ex]} \BibitemShut {NoStop}%
\bibitem [{\citenamefont {Chatrchyan}\ \emph {et~al.}(2012)\citenamefont
  {Chatrchyan} \emph {et~al.}}]{Chatrchyan:2012xdj}%
  \BibitemOpen
  \bibfield  {author} {\bibinfo {author} {\bibfnamefont {S.}~\bibnamefont
  {Chatrchyan}} \emph {et~al.} (\bibinfo {collaboration} {CMS}),\ }\Doi
  {10.1016/j.physletb.2012.08.021} {\bibfield  {journal} {\bibinfo  {journal}
  {Phys. Lett.},\ }\textbf {\bibinfo {volume} {B716}},\ \bibinfo {pages} {30}
  (\bibinfo {year} {2012})},\ \Eprint {http://arxiv.org/abs/1207.7235}
  {arXiv:1207.7235 [hep-ex]} \BibitemShut {NoStop}%
\bibitem [{\citenamefont {Higgs}(1964){\natexlab{a}}}]{Higgs:1964ia}%
  \BibitemOpen
  \bibfield  {author} {\bibinfo {author} {\bibfnamefont {P.~W.}\ \bibnamefont
  {Higgs}},\ }\Doi {10.1016/0031-9163(64)91136-9} {\bibfield  {journal}
  {\bibinfo  {journal} {Phys.Lett.},\ }\textbf {\bibinfo {volume} {12}},\
  \bibinfo {pages} {132} (\bibinfo {year} {1964}{\natexlab{a}})}\BibitemShut
  {NoStop}%
\bibitem [{\citenamefont {Englert}\ and\ \citenamefont
  {Brout}(1964)}]{Englert:1964et}%
  \BibitemOpen
  \bibfield  {author} {\bibinfo {author} {\bibfnamefont {F.}~\bibnamefont
  {Englert}}\ and\ \bibinfo {author} {\bibfnamefont {R.}~\bibnamefont
  {Brout}},\ }\Doi {10.1103/PhysRevLett.13.321} {\bibfield  {journal} {\bibinfo
   {journal} {Phys.Rev.Lett.},\ }\textbf {\bibinfo {volume} {13}},\ \bibinfo
  {pages} {321} (\bibinfo {year} {1964})}\BibitemShut {NoStop}%
\bibitem [{\citenamefont {Higgs}(1964){\natexlab{b}}}]{Higgs:1964pj}%
  \BibitemOpen
  \bibfield  {author} {\bibinfo {author} {\bibfnamefont {P.~W.}\ \bibnamefont
  {Higgs}},\ }\Doi {10.1103/PhysRevLett.13.508} {\bibfield  {journal} {\bibinfo
   {journal} {Phys.Rev.Lett.},\ }\textbf {\bibinfo {volume} {13}},\ \bibinfo
  {pages} {508} (\bibinfo {year} {1964}{\natexlab{b}})}\BibitemShut {NoStop}%
\bibitem [{\citenamefont {Guralnik}\ \emph {et~al.}(1964)\citenamefont
  {Guralnik}, \citenamefont {Hagen},\ and\ \citenamefont
  {Kibble}}]{Guralnik:1964eu}%
  \BibitemOpen
  \bibfield  {author} {\bibinfo {author} {\bibfnamefont {G.}~\bibnamefont
  {Guralnik}}, \bibinfo {author} {\bibfnamefont {C.}~\bibnamefont {Hagen}}, \
  and\ \bibinfo {author} {\bibfnamefont {T.}~\bibnamefont {Kibble}},\ }\Doi
  {10.1103/PhysRevLett.13.585} {\bibfield  {journal} {\bibinfo  {journal}
  {Phys.Rev.Lett.},\ }\textbf {\bibinfo {volume} {13}},\ \bibinfo {pages} {585}
  (\bibinfo {year} {1964})}\BibitemShut {NoStop}%
\bibitem [{\citenamefont {Dawson}\ \emph {et~al.}(1998)\citenamefont {Dawson},
  \citenamefont {Dittmaier},\ and\ \citenamefont {Spira}}]{Dawson:1998py}%
  \BibitemOpen
  \bibfield  {author} {\bibinfo {author} {\bibfnamefont {S.}~\bibnamefont
  {Dawson}}, \bibinfo {author} {\bibfnamefont {S.}~\bibnamefont {Dittmaier}}, \
  and\ \bibinfo {author} {\bibfnamefont {M.}~\bibnamefont {Spira}},\ }\Doi
  {10.1103/PhysRevD.58.115012} {\bibfield  {journal} {\bibinfo  {journal}
  {Phys. Rev.},\ }\textbf {\bibinfo {volume} {D58}},\ \bibinfo {pages} {115012}
  (\bibinfo {year} {1998})},\ \Eprint {http://arxiv.org/abs/hep-ph/9805244}
  {arXiv:hep-ph/9805244 [hep-ph]} \BibitemShut {NoStop}%
\bibitem [{\citenamefont {Djouadi}\ \emph {et~al.}(1999)\citenamefont
  {Djouadi}, \citenamefont {Kilian}, \citenamefont {Muhlleitner},\ and\
  \citenamefont {Zerwas}}]{Djouadi:1999rca}%
  \BibitemOpen
  \bibfield  {author} {\bibinfo {author} {\bibfnamefont {A.}~\bibnamefont
  {Djouadi}}, \bibinfo {author} {\bibfnamefont {W.}~\bibnamefont {Kilian}},
  \bibinfo {author} {\bibfnamefont {M.}~\bibnamefont {Muhlleitner}}, \ and\
  \bibinfo {author} {\bibfnamefont {P.~M.}\ \bibnamefont {Zerwas}},\ }\Doi
  {10.1007/s100529900083} {\bibfield  {journal} {\bibinfo  {journal} {Eur.
  Phys. J.},\ }\textbf {\bibinfo {volume} {C10}},\ \bibinfo {pages} {45}
  (\bibinfo {year} {1999})},\ \Eprint {http://arxiv.org/abs/hep-ph/9904287}
  {arXiv:hep-ph/9904287 [hep-ph]} \BibitemShut {NoStop}%
\bibitem [{\citenamefont {Plehn}\ and\ \citenamefont
  {Rauch}(2005)}]{Plehn:2005nk}%
  \BibitemOpen
  \bibfield  {author} {\bibinfo {author} {\bibfnamefont {T.}~\bibnamefont
  {Plehn}}\ and\ \bibinfo {author} {\bibfnamefont {M.}~\bibnamefont {Rauch}},\
  }\Doi {10.1103/PhysRevD.72.053008} {\bibfield  {journal} {\bibinfo  {journal}
  {Phys. Rev.},\ }\textbf {\bibinfo {volume} {D72}},\ \bibinfo {pages} {053008}
  (\bibinfo {year} {2005})},\ \Eprint {http://arxiv.org/abs/hep-ph/0507321}
  {arXiv:hep-ph/0507321 [hep-ph]} \BibitemShut {NoStop}%
\bibitem [{\citenamefont {Binoth}\ \emph {et~al.}(2006)\citenamefont {Binoth},
  \citenamefont {Karg}, \citenamefont {Kauer},\ and\ \citenamefont
  {Ruckl}}]{Binoth:2006ym}%
  \BibitemOpen
  \bibfield  {author} {\bibinfo {author} {\bibfnamefont {T.}~\bibnamefont
  {Binoth}}, \bibinfo {author} {\bibfnamefont {S.}~\bibnamefont {Karg}},
  \bibinfo {author} {\bibfnamefont {N.}~\bibnamefont {Kauer}}, \ and\ \bibinfo
  {author} {\bibfnamefont {R.}~\bibnamefont {Ruckl}},\ }\Doi
  {10.1103/PhysRevD.74.113008} {\bibfield  {journal} {\bibinfo  {journal}
  {Phys. Rev.},\ }\textbf {\bibinfo {volume} {D74}},\ \bibinfo {pages} {113008}
  (\bibinfo {year} {2006})},\ \Eprint {http://arxiv.org/abs/hep-ph/0608057}
  {arXiv:hep-ph/0608057 [hep-ph]} \BibitemShut {NoStop}%
\bibitem [{\citenamefont {Fuks}\ \emph {et~al.}(2016)\citenamefont {Fuks},
  \citenamefont {Kim},\ and\ \citenamefont {Lee}}]{Fuks:2015hna}%
  \BibitemOpen
  \bibfield  {author} {\bibinfo {author} {\bibfnamefont {B.}~\bibnamefont
  {Fuks}}, \bibinfo {author} {\bibfnamefont {J.~H.}\ \bibnamefont {Kim}}, \
  and\ \bibinfo {author} {\bibfnamefont {S.~J.}\ \bibnamefont {Lee}},\ }\Doi
  {10.1103/PhysRevD.93.035026} {\bibfield  {journal} {\bibinfo  {journal}
  {Phys. Rev.},\ }\textbf {\bibinfo {volume} {D93}},\ \bibinfo {pages} {035026}
  (\bibinfo {year} {2016})},\ \Eprint {http://arxiv.org/abs/1510.07697}
  {arXiv:1510.07697 [hep-ph]} \BibitemShut {NoStop}%
\bibitem [{\citenamefont {Eboli}\ \emph {et~al.}(1987)\citenamefont {Eboli},
  \citenamefont {Marques}, \citenamefont {Novaes},\ and\ \citenamefont
  {Natale}}]{Eboli:1987dy}%
  \BibitemOpen
  \bibfield  {author} {\bibinfo {author} {\bibfnamefont {O.~J.~P.}\
  \bibnamefont {Eboli}}, \bibinfo {author} {\bibfnamefont {G.~C.}\ \bibnamefont
  {Marques}}, \bibinfo {author} {\bibfnamefont {S.~F.}\ \bibnamefont {Novaes}},
  \ and\ \bibinfo {author} {\bibfnamefont {A.~A.}\ \bibnamefont {Natale}},\
  }\Doi {10.1016/0370-2693(87)90381-9} {\bibfield  {journal} {\bibinfo
  {journal} {Phys. Lett.},\ }\textbf {\bibinfo {volume} {B197}},\ \bibinfo
  {pages} {269} (\bibinfo {year} {1987})}\BibitemShut {NoStop}%
\bibitem [{\citenamefont {Glover}\ and\ \citenamefont {van~der
  Bij}(1988)}]{Glover:1987nx}%
  \BibitemOpen
  \bibfield  {author} {\bibinfo {author} {\bibfnamefont {E.~W.~N.}\
  \bibnamefont {Glover}}\ and\ \bibinfo {author} {\bibfnamefont {J.~J.}\
  \bibnamefont {van~der Bij}},\ }\Doi {10.1016/0550-3213(88)90083-1} {\bibfield
   {journal} {\bibinfo  {journal} {Nucl. Phys.},\ }\textbf {\bibinfo {volume}
  {B309}},\ \bibinfo {pages} {282} (\bibinfo {year} {1988})}\BibitemShut
  {NoStop}%
\bibitem [{\citenamefont {Dicus}\ \emph {et~al.}(1988)\citenamefont {Dicus},
  \citenamefont {Kao},\ and\ \citenamefont {Willenbrock}}]{Dicus:1987ic}%
  \BibitemOpen
  \bibfield  {author} {\bibinfo {author} {\bibfnamefont {D.~A.}\ \bibnamefont
  {Dicus}}, \bibinfo {author} {\bibfnamefont {C.}~\bibnamefont {Kao}}, \ and\
  \bibinfo {author} {\bibfnamefont {S.~S.~D.}\ \bibnamefont {Willenbrock}},\
  }\Doi {10.1016/0370-2693(88)90202-X} {\bibfield  {journal} {\bibinfo
  {journal} {Phys. Lett.},\ }\textbf {\bibinfo {volume} {B203}},\ \bibinfo
  {pages} {457} (\bibinfo {year} {1988})}\BibitemShut {NoStop}%
\bibitem [{\citenamefont {Plehn}\ \emph {et~al.}(1996)\citenamefont {Plehn},
  \citenamefont {Spira},\ and\ \citenamefont {Zerwas}}]{Plehn:1996wb}%
  \BibitemOpen
  \bibfield  {author} {\bibinfo {author} {\bibfnamefont {T.}~\bibnamefont
  {Plehn}}, \bibinfo {author} {\bibfnamefont {M.}~\bibnamefont {Spira}}, \ and\
  \bibinfo {author} {\bibfnamefont {P.~M.}\ \bibnamefont {Zerwas}},\ }\Doi
  {10.1016/0550-3213(96)00418-X, 10.1016/S0550-3213(98)00406-4} {\bibfield
  {journal} {\bibinfo  {journal} {Nucl. Phys.},\ }\textbf {\bibinfo {volume}
  {B479}},\ \bibinfo {pages} {46} (\bibinfo {year} {1996})},\ \bibinfo {note}
  {[Erratum: Nucl. Phys.B531,655(1998)]},\ \Eprint
  {http://arxiv.org/abs/hep-ph/9603205} {arXiv:hep-ph/9603205 [hep-ph]}
  \BibitemShut {NoStop}%
\bibitem [{\citenamefont {de~Florian}\ and\ \citenamefont
  {Mazzitelli}(2013){\natexlab{a}}}]{deFlorian:2013uza}%
  \BibitemOpen
  \bibfield  {author} {\bibinfo {author} {\bibfnamefont {D.}~\bibnamefont
  {de~Florian}}\ and\ \bibinfo {author} {\bibfnamefont {J.}~\bibnamefont
  {Mazzitelli}},\ }\Doi {10.1016/j.physletb.2013.06.046} {\bibfield  {journal}
  {\bibinfo  {journal} {Phys. Lett.},\ }\textbf {\bibinfo {volume} {B724}},\
  \bibinfo {pages} {306} (\bibinfo {year} {2013}{\natexlab{a}})},\ \Eprint
  {http://arxiv.org/abs/1305.5206} {arXiv:1305.5206 [hep-ph]} \BibitemShut
  {NoStop}%
\bibitem [{\citenamefont {de~Florian}\ and\ \citenamefont
  {Mazzitelli}(2013){\natexlab{b}}}]{deFlorian:2013jea}%
  \BibitemOpen
  \bibfield  {author} {\bibinfo {author} {\bibfnamefont {D.}~\bibnamefont
  {de~Florian}}\ and\ \bibinfo {author} {\bibfnamefont {J.}~\bibnamefont
  {Mazzitelli}},\ }\Doi {10.1103/PhysRevLett.111.201801} {\bibfield  {journal}
  {\bibinfo  {journal} {Phys. Rev. Lett.},\ }\textbf {\bibinfo {volume}
  {111}},\ \bibinfo {pages} {201801} (\bibinfo {year} {2013}{\natexlab{b}})},\
  \Eprint {http://arxiv.org/abs/1309.6594} {arXiv:1309.6594 [hep-ph]}
  \BibitemShut {NoStop}%
\bibitem [{\citenamefont {Grigo}\ \emph {et~al.}(2014)\citenamefont {Grigo},
  \citenamefont {Melnikov},\ and\ \citenamefont {Steinhauser}}]{Grigo:2014jma}%
  \BibitemOpen
  \bibfield  {author} {\bibinfo {author} {\bibfnamefont {J.}~\bibnamefont
  {Grigo}}, \bibinfo {author} {\bibfnamefont {K.}~\bibnamefont {Melnikov}}, \
  and\ \bibinfo {author} {\bibfnamefont {M.}~\bibnamefont {Steinhauser}},\
  }\Doi {10.1016/j.nuclphysb.2014.09.003} {\bibfield  {journal} {\bibinfo
  {journal} {Nucl. Phys.},\ }\textbf {\bibinfo {volume} {B888}},\ \bibinfo
  {pages} {17} (\bibinfo {year} {2014})},\ \Eprint
  {http://arxiv.org/abs/1408.2422} {arXiv:1408.2422 [hep-ph]} \BibitemShut
  {NoStop}%
\bibitem [{\citenamefont {de~Florian}\ \emph {et~al.}(2016)\citenamefont
  {de~Florian}, \citenamefont {Grazzini}, \citenamefont {Hanga}, \citenamefont
  {Kallweit}, \citenamefont {Lindert}, \citenamefont {{Maierh\"ofer}},
  \citenamefont {Mazzitelli},\ and\ \citenamefont
  {Rathlev}}]{deFlorian:2016uhr}%
  \BibitemOpen
  \bibfield  {author} {\bibinfo {author} {\bibfnamefont {D.}~\bibnamefont
  {de~Florian}}, \bibinfo {author} {\bibfnamefont {M.}~\bibnamefont
  {Grazzini}}, \bibinfo {author} {\bibfnamefont {C.}~\bibnamefont {Hanga}},
  \bibinfo {author} {\bibfnamefont {S.}~\bibnamefont {Kallweit}}, \bibinfo
  {author} {\bibfnamefont {J.~M.}\ \bibnamefont {Lindert}}, \bibinfo {author}
  {\bibfnamefont {P.}~\bibnamefont {{Maierh\"ofer}}}, \bibinfo {author}
  {\bibfnamefont {J.}~\bibnamefont {Mazzitelli}}, \ and\ \bibinfo {author}
  {\bibfnamefont {D.}~\bibnamefont {Rathlev}},\ }\Doi {10.1007/JHEP09(2016)151}
  {\bibfield  {journal} {\bibinfo  {journal} {JHEP},\ }\textbf {\bibinfo
  {volume} {09}},\ \bibinfo {pages} {151} (\bibinfo {year} {2016})},\ \Eprint
  {http://arxiv.org/abs/1606.09519} {arXiv:1606.09519 [hep-ph]} \BibitemShut
  {NoStop}%
\bibitem [{\citenamefont {Shao}\ \emph {et~al.}(2013)\citenamefont {Shao},
  \citenamefont {Li}, \citenamefont {Li},\ and\ \citenamefont
  {Wang}}]{Shao:2013bz}%
  \BibitemOpen
  \bibfield  {author} {\bibinfo {author} {\bibfnamefont {D.~Y.}\ \bibnamefont
  {Shao}}, \bibinfo {author} {\bibfnamefont {C.~S.}\ \bibnamefont {Li}},
  \bibinfo {author} {\bibfnamefont {H.~T.}\ \bibnamefont {Li}}, \ and\ \bibinfo
  {author} {\bibfnamefont {J.}~\bibnamefont {Wang}},\ }\Doi
  {10.1007/JHEP07(2013)169} {\bibfield  {journal} {\bibinfo  {journal} {JHEP},\
  }\textbf {\bibinfo {volume} {07}},\ \bibinfo {pages} {169} (\bibinfo {year}
  {2013})},\ \Eprint {http://arxiv.org/abs/1301.1245} {arXiv:1301.1245
  [hep-ph]} \BibitemShut {NoStop}%
\bibitem [{\citenamefont {de~Florian}\ and\ \citenamefont
  {Mazzitelli}(2015)}]{deFlorian:2015moa}%
  \BibitemOpen
  \bibfield  {author} {\bibinfo {author} {\bibfnamefont {D.}~\bibnamefont
  {de~Florian}}\ and\ \bibinfo {author} {\bibfnamefont {J.}~\bibnamefont
  {Mazzitelli}},\ }\Doi {10.1007/JHEP09(2015)053} {\bibfield  {journal}
  {\bibinfo  {journal} {JHEP},\ }\textbf {\bibinfo {volume} {09}},\ \bibinfo
  {pages} {053} (\bibinfo {year} {2015})},\ \Eprint
  {http://arxiv.org/abs/1505.07122} {arXiv:1505.07122 [hep-ph]} \BibitemShut
  {NoStop}%
\bibitem [{\citenamefont {Spira}(2016)}]{Spira:2016zna}%
  \BibitemOpen
  \bibfield  {author} {\bibinfo {author} {\bibfnamefont {M.}~\bibnamefont
  {Spira}},\ }\Doi {10.1007/JHEP10(2016)026} {\bibfield  {journal} {\bibinfo
  {journal} {JHEP},\ }\textbf {\bibinfo {volume} {10}},\ \bibinfo {pages} {026}
  (\bibinfo {year} {2016})},\ \Eprint {http://arxiv.org/abs/1607.05548}
  {arXiv:1607.05548 [hep-ph]} \BibitemShut {NoStop}%
\bibitem [{\citenamefont {Frederix}\ \emph {et~al.}(2014)\citenamefont
  {Frederix}, \citenamefont {Frixione}, \citenamefont {Hirschi}, \citenamefont
  {Maltoni}, \citenamefont {Mattelaer}, \citenamefont {Torrielli},
  \citenamefont {Vryonidou},\ and\ \citenamefont {Zaro}}]{Frederix:2014hta}%
  \BibitemOpen
  \bibfield  {author} {\bibinfo {author} {\bibfnamefont {R.}~\bibnamefont
  {Frederix}}, \bibinfo {author} {\bibfnamefont {S.}~\bibnamefont {Frixione}},
  \bibinfo {author} {\bibfnamefont {V.}~\bibnamefont {Hirschi}}, \bibinfo
  {author} {\bibfnamefont {F.}~\bibnamefont {Maltoni}}, \bibinfo {author}
  {\bibfnamefont {O.}~\bibnamefont {Mattelaer}}, \bibinfo {author}
  {\bibfnamefont {P.}~\bibnamefont {Torrielli}}, \bibinfo {author}
  {\bibfnamefont {E.}~\bibnamefont {Vryonidou}}, \ and\ \bibinfo {author}
  {\bibfnamefont {M.}~\bibnamefont {Zaro}},\ }\Doi
  {10.1016/j.physletb.2014.03.026} {\bibfield  {journal} {\bibinfo  {journal}
  {Phys. Lett.},\ }\textbf {\bibinfo {volume} {B732}},\ \bibinfo {pages} {142}
  (\bibinfo {year} {2014})},\ \Eprint {http://arxiv.org/abs/1401.7340}
  {arXiv:1401.7340 [hep-ph]} \BibitemShut {NoStop}%
\bibitem [{\citenamefont {Maltoni}\ \emph {et~al.}(2014)\citenamefont
  {Maltoni}, \citenamefont {Vryonidou},\ and\ \citenamefont
  {Zaro}}]{Maltoni:2014eza}%
  \BibitemOpen
  \bibfield  {author} {\bibinfo {author} {\bibfnamefont {F.}~\bibnamefont
  {Maltoni}}, \bibinfo {author} {\bibfnamefont {E.}~\bibnamefont {Vryonidou}},
  \ and\ \bibinfo {author} {\bibfnamefont {M.}~\bibnamefont {Zaro}},\ }\Doi
  {10.1007/JHEP11(2014)079} {\bibfield  {journal} {\bibinfo  {journal} {JHEP},\
  }\textbf {\bibinfo {volume} {11}},\ \bibinfo {pages} {079} (\bibinfo {year}
  {2014})},\ \Eprint {http://arxiv.org/abs/1408.6542} {arXiv:1408.6542
  [hep-ph]} \BibitemShut {NoStop}%
\bibitem [{\citenamefont {Grigo}\ \emph {et~al.}(2013)\citenamefont {Grigo},
  \citenamefont {Hoff}, \citenamefont {Melnikov},\ and\ \citenamefont
  {Steinhauser}}]{Grigo:2013rya}%
  \BibitemOpen
  \bibfield  {author} {\bibinfo {author} {\bibfnamefont {J.}~\bibnamefont
  {Grigo}}, \bibinfo {author} {\bibfnamefont {J.}~\bibnamefont {Hoff}},
  \bibinfo {author} {\bibfnamefont {K.}~\bibnamefont {Melnikov}}, \ and\
  \bibinfo {author} {\bibfnamefont {M.}~\bibnamefont {Steinhauser}},\ }\Doi
  {10.1016/j.nuclphysb.2013.06.024} {\bibfield  {journal} {\bibinfo  {journal}
  {Nucl. Phys.},\ }\textbf {\bibinfo {volume} {B875}},\ \bibinfo {pages} {1}
  (\bibinfo {year} {2013})},\ \Eprint {http://arxiv.org/abs/1305.7340}
  {arXiv:1305.7340 [hep-ph]} \BibitemShut {NoStop}%
\bibitem [{\citenamefont {Grigo}\ \emph {et~al.}(2015)\citenamefont {Grigo},
  \citenamefont {Hoff},\ and\ \citenamefont {Steinhauser}}]{Grigo:2015dia}%
  \BibitemOpen
  \bibfield  {author} {\bibinfo {author} {\bibfnamefont {J.}~\bibnamefont
  {Grigo}}, \bibinfo {author} {\bibfnamefont {J.}~\bibnamefont {Hoff}}, \ and\
  \bibinfo {author} {\bibfnamefont {M.}~\bibnamefont {Steinhauser}},\ }\Doi
  {10.1016/j.nuclphysb.2015.09.012} {\bibfield  {journal} {\bibinfo  {journal}
  {Nucl. Phys.},\ }\textbf {\bibinfo {volume} {B900}},\ \bibinfo {pages} {412}
  (\bibinfo {year} {2015})},\ \Eprint {http://arxiv.org/abs/1508.00909}
  {arXiv:1508.00909 [hep-ph]} \BibitemShut {NoStop}%
\bibitem [{\citenamefont {Borowka}\ \emph
  {et~al.}(2016){\natexlab{a}}\citenamefont {Borowka}, \citenamefont {Greiner},
  \citenamefont {Heinrich}, \citenamefont {Jones}, \citenamefont {Kerner},
  \citenamefont {Schlenk}, \citenamefont {Schubert},\ and\ \citenamefont
  {Zirke}}]{Borowka:2016ehy}%
  \BibitemOpen
  \bibfield  {author} {\bibinfo {author} {\bibfnamefont {S.}~\bibnamefont
  {Borowka}}, \bibinfo {author} {\bibfnamefont {N.}~\bibnamefont {Greiner}},
  \bibinfo {author} {\bibfnamefont {G.}~\bibnamefont {Heinrich}}, \bibinfo
  {author} {\bibfnamefont {S.~P.}\ \bibnamefont {Jones}}, \bibinfo {author}
  {\bibfnamefont {M.}~\bibnamefont {Kerner}}, \bibinfo {author} {\bibfnamefont
  {J.}~\bibnamefont {Schlenk}}, \bibinfo {author} {\bibfnamefont
  {U.}~\bibnamefont {Schubert}}, \ and\ \bibinfo {author} {\bibfnamefont
  {T.}~\bibnamefont {Zirke}},\ }\Doi {10.1103/PhysRevLett.117.079901,
  10.1103/PhysRevLett.117.012001} {\bibfield  {journal} {\bibinfo  {journal}
  {Phys. Rev. Lett.},\ }\textbf {\bibinfo {volume} {117}},\ \bibinfo {pages}
  {012001} (\bibinfo {year} {2016}{\natexlab{a}})},\ \bibinfo {note} {[Erratum:
  Phys. Rev. Lett.117,no.7,079901(2016)]}, \mbox{\Eprint
  {http://arxiv.org/abs/1604.06447} {arXiv:1604.06447 [hep-ph]}} \BibitemShut
  {NoStop}%
\bibitem [{\citenamefont {Borowka}\ \emph
  {et~al.}(2016){\natexlab{b}}\citenamefont {Borowka}, \citenamefont {Greiner},
  \citenamefont {Heinrich}, \citenamefont {Jones}, \citenamefont {Kerner},
  \citenamefont {Schlenk},\ and\ \citenamefont {Zirke}}]{Borowka:2016ypz}%
  \BibitemOpen
  \bibfield  {author} {\bibinfo {author} {\bibfnamefont {S.}~\bibnamefont
  {Borowka}}, \bibinfo {author} {\bibfnamefont {N.}~\bibnamefont {Greiner}},
  \bibinfo {author} {\bibfnamefont {G.}~\bibnamefont {Heinrich}}, \bibinfo
  {author} {\bibfnamefont {S.~P.}\ \bibnamefont {Jones}}, \bibinfo {author}
  {\bibfnamefont {M.}~\bibnamefont {Kerner}}, \bibinfo {author} {\bibfnamefont
  {J.}~\bibnamefont {Schlenk}}, \ and\ \bibinfo {author} {\bibfnamefont
  {T.}~\bibnamefont {Zirke}},\ }\Doi {10.1007/JHEP10(2016)107} {\bibfield
  {journal} {\bibinfo  {journal} {JHEP},\ }\textbf {\bibinfo {volume} {10}},\
  \bibinfo {pages} {107} (\bibinfo {year} {2016}{\natexlab{b}})},\ \Eprint
  {http://arxiv.org/abs/1608.04798} {arXiv:1608.04798 [hep-ph]} \BibitemShut
  {NoStop}%
\bibitem [{\citenamefont {Heinrich}\ \emph {et~al.}(2017)\citenamefont
  {Heinrich}, \citenamefont {Jones}, \citenamefont {Kerner}, \citenamefont
  {Luisoni},\ and\ \citenamefont {Vryonidou}}]{Heinrich:2017kxx}%
  \BibitemOpen
  \bibfield  {author} {\bibinfo {author} {\bibfnamefont {G.}~\bibnamefont
  {Heinrich}}, \bibinfo {author} {\bibfnamefont {S.~P.}\ \bibnamefont {Jones}},
  \bibinfo {author} {\bibfnamefont {M.}~\bibnamefont {Kerner}}, \bibinfo
  {author} {\bibfnamefont {G.}~\bibnamefont {Luisoni}}, \ and\ \bibinfo
  {author} {\bibfnamefont {E.}~\bibnamefont {Vryonidou}},\ }\Doi
  {10.1007/JHEP08(2017)088} {\bibfield  {journal} {\bibinfo  {journal} {JHEP},\
  }\textbf {\bibinfo {volume} {08}},\ \bibinfo {pages} {088} (\bibinfo {year}
  {2017})},\ \Eprint {http://arxiv.org/abs/1703.09252} {arXiv:1703.09252
  [hep-ph]} \BibitemShut {NoStop}%
\bibitem [{\citenamefont {Jones}\ and\ \citenamefont
  {Kuttimalai}(2018)}]{Jones:2017giv}%
  \BibitemOpen
  \bibfield  {author} {\bibinfo {author} {\bibfnamefont {S.}~\bibnamefont
  {Jones}}\ and\ \bibinfo {author} {\bibfnamefont {S.}~\bibnamefont
  {Kuttimalai}},\ }\Doi {10.1007/JHEP02(2018)176} {\bibfield  {journal}
  {\bibinfo  {journal} {JHEP},\ }\textbf {\bibinfo {volume} {02}},\ \bibinfo
  {pages} {176} (\bibinfo {year} {2018})},\ \Eprint
  {http://arxiv.org/abs/1711.03319} {arXiv:1711.03319 [hep-ph]} \BibitemShut
  {NoStop}%
\bibitem [{\citenamefont {Grazzini}\ \emph {et~al.}(2018)\citenamefont
  {Grazzini}, \citenamefont {Heinrich}, \citenamefont {Jones}, \citenamefont
  {Kallweit}, \citenamefont {Kerner}, \citenamefont {Lindert},\ and\
  \citenamefont {Mazzitelli}}]{Grazzini:2018bsd}%
  \BibitemOpen
  \bibfield  {author} {\bibinfo {author} {\bibfnamefont {M.}~\bibnamefont
  {Grazzini}}, \bibinfo {author} {\bibfnamefont {G.}~\bibnamefont {Heinrich}},
  \bibinfo {author} {\bibfnamefont {S.}~\bibnamefont {Jones}}, \bibinfo
  {author} {\bibfnamefont {S.}~\bibnamefont {Kallweit}}, \bibinfo {author}
  {\bibfnamefont {M.}~\bibnamefont {Kerner}}, \bibinfo {author} {\bibfnamefont
  {J.~M.}\ \bibnamefont {Lindert}}, \ and\ \bibinfo {author} {\bibfnamefont
  {J.}~\bibnamefont {Mazzitelli}},\ }\Doi {10.1007/JHEP05(2018)059} {\bibfield
  {journal} {\bibinfo  {journal} {JHEP},\ }\textbf {\bibinfo {volume} {05}},\
  \bibinfo {pages} {059} (\bibinfo {year} {2018})},\ \Eprint
  {http://arxiv.org/abs/1803.02463} {arXiv:1803.02463 [hep-ph]} \BibitemShut
  {NoStop}%
\bibitem [{\citenamefont {{Gr\"ober}}\ \emph {et~al.}(2018)\citenamefont
  {{Gr\"ober}}, \citenamefont {Maier},\ and\ \citenamefont
  {Rauh}}]{Grober:2017uho}%
  \BibitemOpen
  \bibfield  {author} {\bibinfo {author} {\bibfnamefont {R.}~\bibnamefont
  {{Gr\"ober}}}, \bibinfo {author} {\bibfnamefont {A.}~\bibnamefont {Maier}}, \
  and\ \bibinfo {author} {\bibfnamefont {T.}~\bibnamefont {Rauh}},\ }\Doi
  {10.1007/JHEP03(2018)020} {\bibfield  {journal} {\bibinfo  {journal} {JHEP},\
  }\textbf {\bibinfo {volume} {03}},\ \bibinfo {pages} {020} (\bibinfo {year}
  {2018})},\ \Eprint {http://arxiv.org/abs/1709.07799} {arXiv:1709.07799
  [hep-ph]} \BibitemShut {NoStop}%
\bibitem [{\citenamefont {Bonciani}\ \emph {et~al.}(2018)\citenamefont
  {Bonciani}, \citenamefont {Degrassi}, \citenamefont {Giardino},\ and\
  \citenamefont {{Gr\"ober}}}]{Bonciani:2018omm}%
  \BibitemOpen
  \bibfield  {author} {\bibinfo {author} {\bibfnamefont {R.}~\bibnamefont
  {Bonciani}}, \bibinfo {author} {\bibfnamefont {G.}~\bibnamefont {Degrassi}},
  \bibinfo {author} {\bibfnamefont {P.~P.}\ \bibnamefont {Giardino}}, \ and\
  \bibinfo {author} {\bibfnamefont {R.}~\bibnamefont {{Gr\"ober}}},\ }\Doi
  {10.1103/PhysRevLett.121.162003} {\bibfield  {journal} {\bibinfo  {journal}
  {Phys. Rev. Lett.},\ }\textbf {\bibinfo {volume} {121}},\ \bibinfo {pages}
  {162003} (\bibinfo {year} {2018})},\ \Eprint
  {http://arxiv.org/abs/1806.11564} {arXiv:1806.11564 [hep-ph]} \BibitemShut
  {NoStop}%
\bibitem [{\citenamefont {Davies}\ \emph {et~al.}(2018)\citenamefont {Davies},
  \citenamefont {Mishima}, \citenamefont {Steinhauser},\ and\ \citenamefont
  {Wellmann}}]{Davies:2018qvx}%
  \BibitemOpen
  \bibfield  {author} {\bibinfo {author} {\bibfnamefont {J.}~\bibnamefont
  {Davies}}, \bibinfo {author} {\bibfnamefont {G.}~\bibnamefont {Mishima}},
  \bibinfo {author} {\bibfnamefont {M.}~\bibnamefont {Steinhauser}}, \ and\
  \bibinfo {author} {\bibfnamefont {D.}~\bibnamefont {Wellmann}},\ }
  \Doi {10.1007/JHEP01(2019)176} {\bibfield  {journal}
  {\bibinfo  {journal} {JHEP},\ }\textbf {\bibinfo {volume} {01}},\ \bibinfo
  {pages} {176} (\bibinfo {year} {2019})},\ \Eprint
  {http://arxiv.org/abs/1811.05489} {arXiv:1811.05489 [hep-ph]} \BibitemShut
  {NoStop}%
\bibitem [{\citenamefont {Richardson}(1911)}]{richardson}%
  \BibitemOpen
  \bibfield  {author} {\bibinfo {author} {\bibfnamefont {L.~F.}\ \bibnamefont
  {Richardson}},\ }\Doi {10.1098/rsta.1911.0009} {\bibfield  {journal}
  {\bibinfo  {journal} {Philosophical Transactions of the Royal Society},\
  }\textbf {\bibinfo {volume} {A210}},\ \bibinfo {pages} {307} (\bibinfo {year}
  {1911})}\BibitemShut {NoStop}%
\bibitem [{\citenamefont {Altarelli}\ and\ \citenamefont
  {Parisi}(1977)}]{Altarelli:1977zs}%
  \BibitemOpen
  \bibfield  {author} {\bibinfo {author} {\bibfnamefont {G.}~\bibnamefont
  {Altarelli}}\ and\ \bibinfo {author} {\bibfnamefont {G.}~\bibnamefont
  {Parisi}},\ }\Doi {10.1016/0550-3213(77)90384-4} {\bibfield  {journal}
  {\bibinfo  {journal} {Nucl. Phys.},\ }\textbf {\bibinfo {volume} {B126}},\
  \bibinfo {pages} {298} (\bibinfo {year} {1977})}\BibitemShut {NoStop}%
\bibitem [{\citenamefont {Graudenz}\ \emph {et~al.}(1993)\citenamefont
  {Graudenz}, \citenamefont {Spira},\ and\ \citenamefont
  {Zerwas}}]{Graudenz:1992pv}%
  \BibitemOpen
  \bibfield  {author} {\bibinfo {author} {\bibfnamefont {D.}~\bibnamefont
  {Graudenz}}, \bibinfo {author} {\bibfnamefont {M.}~\bibnamefont {Spira}}, \
  and\ \bibinfo {author} {\bibfnamefont {P.~M.}\ \bibnamefont {Zerwas}},\ }\Doi
  {10.1103/PhysRevLett.70.1372} {\bibfield  {journal} {\bibinfo  {journal}
  {Phys. Rev. Lett.},\ }\textbf {\bibinfo {volume} {70}},\ \bibinfo {pages}
  {1372} (\bibinfo {year} {1993})}\BibitemShut {NoStop}%
\bibitem [{\citenamefont {Spira}\ \emph {et~al.}(1995)\citenamefont {Spira},
  \citenamefont {Djouadi}, \citenamefont {Graudenz},\ and\ \citenamefont
  {Zerwas}}]{Spira:1995rr}%
  \BibitemOpen
  \bibfield  {author} {\bibinfo {author} {\bibfnamefont {M.}~\bibnamefont
  {Spira}}, \bibinfo {author} {\bibfnamefont {A.}~\bibnamefont {Djouadi}},
  \bibinfo {author} {\bibfnamefont {D.}~\bibnamefont {Graudenz}}, \ and\
  \bibinfo {author} {\bibfnamefont {P.~M.}\ \bibnamefont {Zerwas}},\ }\Doi
  {10.1016/0550-3213(95)00379-7} {\bibfield  {journal} {\bibinfo  {journal}
  {Nucl. Phys.},\ }\textbf {\bibinfo {volume} {B453}},\ \bibinfo {pages} {17}
  (\bibinfo {year} {1995})},\ \Eprint {http://arxiv.org/abs/hep-ph/9504378}
  {arXiv:hep-ph/9504378 [hep-ph]} \BibitemShut {NoStop}%
\bibitem [{\citenamefont {Harlander}\ and\ \citenamefont
  {Kant}(2005)}]{Harlander:2005rq}%
  \BibitemOpen
  \bibfield  {author} {\bibinfo {author} {\bibfnamefont {R.}~\bibnamefont
  {Harlander}}\ and\ \bibinfo {author} {\bibfnamefont {P.}~\bibnamefont
  {Kant}},\ }\Doi {10.1088/1126-6708/2005/12/015} {\bibfield  {journal}
  {\bibinfo  {journal} {JHEP},\ }\textbf {\bibinfo {volume} {12}},\ \bibinfo
  {pages} {015} (\bibinfo {year} {2005})},\ \Eprint
  {http://arxiv.org/abs/hep-ph/0509189} {arXiv:hep-ph/0509189 [hep-ph]}
  \BibitemShut {NoStop}%
\bibitem [{\citenamefont {Anastasiou}\ \emph {et~al.}(2009)\citenamefont
  {Anastasiou}, \citenamefont {Bucherer},\ and\ \citenamefont
  {Kunszt}}]{Anastasiou:2009kn}%
  \BibitemOpen
  \bibfield  {author} {\bibinfo {author} {\bibfnamefont {C.}~\bibnamefont
  {Anastasiou}}, \bibinfo {author} {\bibfnamefont {S.}~\bibnamefont
  {Bucherer}}, \ and\ \bibinfo {author} {\bibfnamefont {Z.}~\bibnamefont
  {Kunszt}},\ }\Doi {10.1088/1126-6708/2009/10/068} {\bibfield  {journal}
  {\bibinfo  {journal} {JHEP},\ }\textbf {\bibinfo {volume} {10}},\ \bibinfo
  {pages} {068} (\bibinfo {year} {2009})},\ \Eprint
  {http://arxiv.org/abs/0907.2362} {arXiv:0907.2362 [hep-ph]} \BibitemShut
  {NoStop}%
\bibitem [{\citenamefont {Aglietti}\ \emph {et~al.}(2007)\citenamefont
  {Aglietti}, \citenamefont {Bonciani}, \citenamefont {Degrassi},\ and\
  \citenamefont {Vicini}}]{Aglietti:2006tp}%
  \BibitemOpen
  \bibfield  {author} {\bibinfo {author} {\bibfnamefont {U.}~\bibnamefont
  {Aglietti}}, \bibinfo {author} {\bibfnamefont {R.}~\bibnamefont {Bonciani}},
  \bibinfo {author} {\bibfnamefont {G.}~\bibnamefont {Degrassi}}, \ and\
  \bibinfo {author} {\bibfnamefont {A.}~\bibnamefont {Vicini}},\ }\Doi
  {10.1088/1126-6708/2007/01/021} {\bibfield  {journal} {\bibinfo  {journal}
  {JHEP},\ }\textbf {\bibinfo {volume} {01}},\ \bibinfo {pages} {021} (\bibinfo
  {year} {2007})},\ \Eprint {http://arxiv.org/abs/hep-ph/0611266}
  {arXiv:hep-ph/0611266 [hep-ph]} \BibitemShut {NoStop}%
\bibitem [{\citenamefont {Cahn}\ \emph {et~al.}(1979)\citenamefont {Cahn},
  \citenamefont {Chanowitz},\ and\ \citenamefont {Fleishon}}]{Cahn:1978nz}%
  \BibitemOpen
  \bibfield  {author} {\bibinfo {author} {\bibfnamefont {R.~N.}\ \bibnamefont
  {Cahn}}, \bibinfo {author} {\bibfnamefont {M.~S.}\ \bibnamefont {Chanowitz}},
  \ and\ \bibinfo {author} {\bibfnamefont {N.}~\bibnamefont {Fleishon}},\ }\Doi
  {10.1016/0370-2693(79)90438-6} {\bibfield  {journal} {\bibinfo  {journal}
  {Phys. Lett.},\ }\textbf {\bibinfo {volume} {82B}},\ \bibinfo {pages} {113}
  (\bibinfo {year} {1979})}\BibitemShut {NoStop}%
\bibitem [{\citenamefont {Bergstrom}\ and\ \citenamefont
  {Hulth}(1985)}]{Bergstrom:1985hp}%
  \BibitemOpen
  \bibfield  {author} {\bibinfo {author} {\bibfnamefont {L.}~\bibnamefont
  {Bergstrom}}\ and\ \bibinfo {author} {\bibfnamefont {G.}~\bibnamefont
  {Hulth}},\ }\Doi {10.1016/0550-3213(86)90074-X, 10.1016/0550-3213(85)90302-5}
  {\bibfield  {journal} {\bibinfo  {journal} {Nucl. Phys.},\ }\textbf {\bibinfo
  {volume} {B259}},\ \bibinfo {pages} {137} (\bibinfo {year} {1985})},\
  \bibinfo {note} {[Erratum: Nucl. Phys.B276,744(1986)]}\BibitemShut {NoStop}%
\bibitem [{\citenamefont {Degrassi}\ \emph {et~al.}(2016)\citenamefont
  {Degrassi}, \citenamefont {Giardino},\ and\ \citenamefont
  {{Gr{\"o}ber}}}]{Degrassi:2016vss}%
  \BibitemOpen
  \bibfield  {author} {\bibinfo {author} {\bibfnamefont {G.}~\bibnamefont
  {Degrassi}}, \bibinfo {author} {\bibfnamefont {P.~P.}\ \bibnamefont
  {Giardino}}, \ and\ \bibinfo {author} {\bibfnamefont {R.}~\bibnamefont
  {{Gr{\"o}ber}}},\ }\Doi {10.1140/epjc/s10052-016-4256-9} {\bibfield
  {journal} {\bibinfo  {journal} {Eur. Phys. J.},\ }\textbf {\bibinfo {volume}
  {C76}},\ \bibinfo {pages} {411} (\bibinfo {year} {2016})},\ \Eprint
  {http://arxiv.org/abs/1603.00385} {arXiv:1603.00385 [hep-ph]} \BibitemShut
  {NoStop}%
\bibitem [{\citenamefont {Catani}\ and\ \citenamefont
  {Seymour}(1997)}]{Catani:1996vz}%
  \BibitemOpen
  \bibfield  {author} {\bibinfo {author} {\bibfnamefont {S.}~\bibnamefont
  {Catani}}\ and\ \bibinfo {author} {\bibfnamefont {M.~H.}\ \bibnamefont
  {Seymour}},\ }\Doi {10.1016/S0550-3213(96)00589-5,
  10.1016/S0550-3213(98)81022-5} {\bibfield  {journal} {\bibinfo  {journal}
  {Nucl. Phys.},\ }\textbf {\bibinfo {volume} {B485}},\ \bibinfo {pages} {291}
  (\bibinfo {year} {1997})},\ \bibinfo {note} {[Erratum: Nucl.
  Phys.B510,503(1998)]},\ \Eprint {http://arxiv.org/abs/hep-ph/9605323}
  {arXiv:hep-ph/9605323 [hep-ph]} \BibitemShut {NoStop}%
\bibitem [{\citenamefont {Hahn}(2001)}]{Hahn:2000kx}%
  \BibitemOpen
  \bibfield  {author} {\bibinfo {author} {\bibfnamefont {T.}~\bibnamefont
  {Hahn}},\ }\Doi {10.1016/S0010-4655(01)00290-9} {\bibfield  {journal}
  {\bibinfo  {journal} {Comput. Phys. Commun.},\ }\textbf {\bibinfo {volume}
  {140}},\ \bibinfo {pages} {418} (\bibinfo {year} {2001})},\ \Eprint
  {http://arxiv.org/abs/hep-ph/0012260} {arXiv:hep-ph/0012260 [hep-ph]}
  \BibitemShut {NoStop}%
\bibitem [{\citenamefont {Hahn}\ and\ \citenamefont
  {Perez-Victoria}(1999)}]{Hahn:1998yk}%
  \BibitemOpen
  \bibfield  {author} {\bibinfo {author} {\bibfnamefont {T.}~\bibnamefont
  {Hahn}}\ and\ \bibinfo {author} {\bibfnamefont {M.}~\bibnamefont
  {Perez-Victoria}},\ }\Doi {10.1016/S0010-4655(98)00173-8} {\bibfield
  {journal} {\bibinfo  {journal} {Comput. Phys. Commun.},\ }\textbf {\bibinfo
  {volume} {118}},\ \bibinfo {pages} {153} (\bibinfo {year} {1999})},\ \Eprint
  {http://arxiv.org/abs/hep-ph/9807565} {arXiv:hep-ph/9807565 [hep-ph]}
  \BibitemShut {NoStop}%
\bibitem [{\citenamefont {Denner}\ \emph {et~al.}(2017)\citenamefont {Denner},
  \citenamefont {Dittmaier},\ and\ \citenamefont {Hofer}}]{Denner:2016kdg}%
  \BibitemOpen
  \bibfield  {author} {\bibinfo {author} {\bibfnamefont {A.}~\bibnamefont
  {Denner}}, \bibinfo {author} {\bibfnamefont {S.}~\bibnamefont {Dittmaier}}, \
  and\ \bibinfo {author} {\bibfnamefont {L.}~\bibnamefont {Hofer}},\ }\Doi
  {10.1016/j.cpc.2016.10.013} {\bibfield  {journal} {\bibinfo  {journal}
  {Comput. Phys. Commun.},\ }\textbf {\bibinfo {volume} {212}},\ \bibinfo
  {pages} {220} (\bibinfo {year} {2017})},\ \Eprint
  {http://arxiv.org/abs/1604.06792} {arXiv:1604.06792 [hep-ph]} \BibitemShut
  {NoStop}%
\bibitem [{\citenamefont {Harland-Lang}\ \emph {et~al.}(2015)\citenamefont
  {Harland-Lang}, \citenamefont {Martin}, \citenamefont {Motylinski},\ and\
  \citenamefont {Thorne}}]{Harland-Lang:2014zoa}%
  \BibitemOpen
  \bibfield  {author} {\bibinfo {author} {\bibfnamefont {L.~A.}\ \bibnamefont
  {Harland-Lang}}, \bibinfo {author} {\bibfnamefont {A.~D.}\ \bibnamefont
  {Martin}}, \bibinfo {author} {\bibfnamefont {P.}~\bibnamefont {Motylinski}},
  \ and\ \bibinfo {author} {\bibfnamefont {R.~S.}\ \bibnamefont {Thorne}},\
  }\Doi {10.1140/epjc/s10052-015-3397-6} {\bibfield  {journal} {\bibinfo
  {journal} {Eur. Phys. J.},\ }\textbf {\bibinfo {volume} {C75}},\ \bibinfo
  {pages} {204} (\bibinfo {year} {2015})},\ \ \ \Eprint
  {http://arxiv.org/abs/1412.3989} {arXiv:1412.3989 [hep-ph]} \BibitemShut
  {NoStop}%
\bibitem [{\citenamefont {Butterworth}\ \emph {et~al.}(2016)\citenamefont
  {Butterworth} \emph {et~al.}}]{Butterworth:2015oua}%
  \BibitemOpen
  \bibfield  {author} {\bibinfo {author} {\bibfnamefont {J.}~\bibnamefont
  {Butterworth}} \emph {et~al.},\ }\Doi {10.1088/0954-3899/43/2/023001}
  {\bibfield  {journal} {\bibinfo  {journal} {J. Phys.},\ }\textbf {\bibinfo
  {volume} {G43}},\ \bibinfo {pages} {023001} (\bibinfo {year} {2016})},\
  \Eprint {http://arxiv.org/abs/1510.03865} {arXiv:1510.03865 [hep-ph]}
  \BibitemShut {NoStop}%
\bibitem [{\citenamefont {Buckley}\ \emph {et~al.}(2015)\citenamefont
  {Buckley}, \citenamefont {Ferrando}, \citenamefont {Lloyd}, \citenamefont
  {{Nordstr\"om}}, \citenamefont {Page}, \citenamefont {{R\"ufenacht}},
  \citenamefont {{Sch\"onherr}},\ and\ \citenamefont {Watt}}]{Buckley:2014ana}%
  \BibitemOpen
  \bibfield  {author} {\bibinfo {author} {\bibfnamefont {A.}~\bibnamefont
  {Buckley}}, \bibinfo {author} {\bibfnamefont {J.}~\bibnamefont {Ferrando}},
  \bibinfo {author} {\bibfnamefont {S.}~\bibnamefont {Lloyd}}, \bibinfo
  {author} {\bibfnamefont {K.}~\bibnamefont {{Nordstr\"om}}}, \bibinfo {author}
  {\bibfnamefont {B.}~\bibnamefont {Page}}, \bibinfo {author} {\bibfnamefont
  {M.}~\bibnamefont {{R\"ufenacht}}}, \bibinfo {author} {\bibfnamefont
  {M.}~\bibnamefont {{Sch\"onherr}}}, \ and\ \bibinfo {author} {\bibfnamefont
  {G.}~\bibnamefont {Watt}},\ }\Doi {10.1140/epjc/s10052-015-3318-8} {\bibfield
   {journal} {\bibinfo  {journal} {Eur. Phys. J.},\ }\textbf {\bibinfo {volume}
  {C75}},\ \bibinfo {pages} {132} (\bibinfo {year} {2015})},\ \Eprint
  {http://arxiv.org/abs/1412.7420} {arXiv:1412.7420 [hep-ph]} \BibitemShut
  {NoStop}%
\bibitem [{\citenamefont {Abramowitz}\ and\ \citenamefont
  {Stegun}(1964)}]{Abramowitz}%
  \BibitemOpen
  \bibfield  {author} {\bibinfo {author} {\bibfnamefont {M.}~\bibnamefont
  {Abramowitz}}\ and\ \bibinfo {author} {\bibfnamefont {I.}~\bibnamefont
  {Stegun}},\ }\href@noop {} {\bibfield  {journal} {\bibinfo  {journal}
  {Applied Mathematics Series},\ }\textbf {\bibinfo {volume} {55}},\ \bibinfo
  {pages} {307} (\bibinfo {year} {1964})}\BibitemShut {NoStop}%
\bibitem [{\citenamefont {Baglio}\ \emph {et~al.}(2019)\citenamefont {Baglio},
  \citenamefont {Campanario}, \citenamefont {Glaus}, \citenamefont
  {{M\"uhlleitner}}, \citenamefont {Spira},\ and\ \citenamefont
  {Streicher}}]{newpaper}%
  \BibitemOpen
  \bibfield  {author} {\bibinfo {author} {\bibfnamefont {J.}~\bibnamefont
  {Baglio}}, \bibinfo {author} {\bibfnamefont {F.}~\bibnamefont {Campanario}},
  \bibinfo {author} {\bibfnamefont {S.}~\bibnamefont {Glaus}}, \bibinfo
  {author} {\bibfnamefont {M.}~\bibnamefont {{M\"uhlleitner}}}, \bibinfo
  {author} {\bibfnamefont {M.}~\bibnamefont {Spira}}, \ and\ \bibinfo {author}
  {\bibfnamefont {J.}~\bibnamefont {Streicher}},\ }\href@noop {} {\bibfield
  {journal} {\bibinfo  {journal} {in preparation}} (\bibinfo {year}
  {2019})}\BibitemShut {NoStop}%
\end{thebibliography}%
}

\end{document}